\documentclass[./bst/sn-mathphys]{sn-jnl}

  \usepackage[authoryear]{natbib}%
\jyear{2021}%

\DeclareTextCommand{\DJ}{OT1}{%
  \raisebox{-0.1ex}{\scalebox{0.75}[1.4]{--}}\kern-.4em D%
  }
  \def \dj{d\kern-0.4em\char"16\kern-0.1em} 

\begin{document}

\title[Article Title]{Asteroid Families: properties, recent advances and future opportunities}


\author*[1]{\fnm{Bojan} \sur{Novakovi\'c}}\email{bojan@matf.bg.ac.rs}

\author[2]{\fnm{David} \sur{Vokrouhlick\'y}}

\author[3]{\fnm{Federica} \sur{Spoto}}

\author[4]{\fnm{David} \sur{Nesvorn\'y}}



\affil*[1]{\orgdiv{Department of Astronomy}, \orgname{Faculty of Mathematics, University of Belgrade}, \orgaddress{\street{Studentski trg 16}, \city{Belgrade}, \postcode{11000}, \country{Serbia}}}

\affil[2]{\orgdiv{Institute of Astronomy}, \orgname{Charles University}, \orgaddress{\street{Hole{\v s}ovi{\v c}k{\' a}ch 2}, \city{Prague}, \postcode{CZ-180 00}, \country{Czech Republic}}}

\affil[3]{\orgname{Harvard-Smithsonian Center for Astrophysics}, \orgaddress{\street{60 Garden St, MS 15}, \city{Cambridge}, \postcode{02138}, \state{MA}, \country{USA}}}

\affil[4]{\orgname{Southwest Research Institute}, \orgaddress{\street{1050 Walnut St, Suite 300}, \city{Boulder}, \postcode{80302}, \state{CO}, \country{USA}}}


\abstract{Collisions are one of the key processes shaping planetary systems. Asteroid families are outcomes of such collision still identifiable across our solar system. The families provide a unique view of catastrophic disruption phenomena and have been in the focus of planetary scientists for more than a century. Most of them are located in the main belt, a ring of asteroids between Mars and Jupiter. Here we review the basic properties of the families, discuss some recent advances, and anticipate future challenges. This review pays more attention to dynamic aspects such as family identification, age determination, and long-term evolution. The text, however, goes beyond that. Especially, we cover the details of young families that see the major advances in the last years, and we anticipate it will develop even faster in the future.
We also discuss the relevance of asteroid families for water-ice content in the asteroid belt and our current knowledge on links between families and main-belt comets.}




\maketitle

\section{Introduction}
\label{sec1}

It was more than a century ago when Japanese astronomer Kiyotsugu Hirayama noticed that some asteroids orbit around the Sun along unusually close trajectories \citep{1918AJ.....31..185H}. Hirayama called them \emph{groups of asteroids probably of common origin}, probably not being fully aware of how important topic in asteroid science he initiated. At this point, a fruitful area of asteroid families related studies has begun.

\emph{Why do scientists care about families?}~Collisions among small solar system bodies are of low probability and therefore unlikely to occur over one's lifetime. However, over the age of the system, many such collisions occurred and played key roles in shaping our planetary system.

The dating of asteroid families provides us with a collisional history of the main belt \citep{2015aste.book..701B,2015Icar..257..275S}. It provides a way to rewind a clock and understand what happened at the beginning of the solar system formation and its subsequent evolution. In addition to that, families provide a window into the interiors of their parent bodies, otherwise not accessible, allowing us to understand better the internal composition and structure of the asteroids \citep{masiero15_asteroidIV}. This, in turn, helps us also study the heterogeneity in the protoplanetary disks and mechanisms of planetesimals formation.

Asteroid families also provide unique insights into the forces shaping our planetary system. Families provide us with observable evidence of large-scale catastrophic impacts, giving us the tools to test impact physics on planetary scales \citep{michel15_asteroidIV}. Furthermore, families have been critical in revealing the fingerprints left by mean motion and secular orbital resonances, as well as the non-gravitational Yarkovsky effect \citep{2015aste.book..297N, 2018P&SS..157...72C}. 

Asteroid collisions create thousands of smaller rocks of different sizes. Since their formation, the family members have been subject to different evolutionary processes, including collisional and dynamical evolution. At least some well-placed families are likely to supply a number of fragments to the near-Earth region, connecting, therefore, these families to changes in the impact flux of inner solar system bodies \citep{2017NatAs...1E..35H}.

Finally, the parent bodies of some families were likely water-bearing asteroids. Following the evolution of the fragments formed by collisional disruption of such bodies allows estimating a present-day water's distribution and its content in asteroids \citep{2018AJ....155...96H,2020AJ....159..179H}, relevant in different aspects, including the water delivery to the Earth. Asteroid families are \emph{natural laboratories} to study all these phenomena.

We aim here to put together the results of recent studies on asteroid families, which help navigate current achievements, open problems, and future opportunities. The article focuses on recent advances, achieved mainly since the year 2015. We have tried to avoid duplication with similar attempts, especially two Asteroids IV chapters on asteroids. Therefore, this review should be considered primarily complementary to these works. However, some overlapping is unavoidable, mostly when recalling the previous results is needed to understand the new works.

The paper is organised as follows. In Section~\ref{s:identification} we discussed the methods for family identification and attribution of new members to the known families. Section~\ref{s:dynevo}
review the efforts on modelling the long-term evolution of asteroid families and related processes. At the same time, Section~\ref{sec:special_families} presents the results on special family classes such as young and very young families and water-bearing families. Each section ends by discussing future challenges and opportunities to study the related open problems on asteroid families. In Section~\ref{sec:conclusions} we briefly present our conclusions.

\section{Identification}
\label{s:identification}

This section reviews the existing methods for classifying asteroids into families and discusses the main challenges that need to be addressed in the future. Methods especially suitable for young families, such as the backward integration method, are discussed in Section~\ref{yf}.

\subsection{Traditional Hierarchical Clustering Method}
\label{ss:clasic_hcm}

A standard approach to identifying asteroid families is the so-called Hierarchical Clustering Method (HCM), introduced by \citet{1990AJ....100.2030Z}. This the most often used methodology is extensively described in the literature. In particular, the advantages and disadvantages of the HCM are discussed, for instance, by \citet{1993A&A...272..651B} or \citet{2015aste.book..297N}. Here we only briefly review the main idea and properties of the method.

Detection of asteroid families and identification of their members is typically performed in the 3-D space of proper orbital elements: semi-major axis $a_{\rm p}$, eccentricity $e_{\rm p}$, and inclination $i_{\rm p}$ \citep[][]{2014Icar..239...46M,2015aste.book..297N}, though in principle this could be done also in the space of proper frequencies \citep[][]{2007A&A...475.1145C,2009A&A...493..267C}.

The distances among the asteroids in the proper elements space are usually determined using the following metrics proposed by \citet{1990AJ....100.2030Z}:
\begin{equation}
 d = na_{\rm p} \sqrt{\dfrac{5}{4} \left( \dfrac{\Delta a_{\rm p}}{a_{\rm p}} \right)^2 + 2(\Delta e_{\rm p})^2 + 2(\Delta \sin i_{\rm p})^2},  
 \label{eq:hcm_dis}
\end{equation}
where $na_{\rm p}$ is the heliocentric orbital velocity of an asteroid on a circular orbit having the semi-major axis $a_{\rm p}$ = ($a_{{\rm p},2}+a_{{\rm p},1}$)/2, $\Delta a_{\rm p} = a_{{\rm p},1} - a_{{\rm p},2}$, $\Delta e_{\rm p} = e_{{\rm p},1} - e_{{\rm p},2}$ and $\Delta \sin i_{\rm p} = \sin i_{{\rm p},1} - \sin i_{{\rm p},2}$, 
where the indexes (1) and (2) denote the two bodies whose mutual distance is calculated.  The distance $d$ is usually expressed in meters per second.

The HCM used to identify asteroid families is based on a single-linkage clustering. It works in an agglomerative (bottom-up) fashion so that at each step, two clusters that contain the closest pair of elements, not yet belonging to the same cluster, are combined. The main advantage of the single-linkage HCM is that it does not preclude apriori any shape of the family. Therefore, it allows the detection of irregular clusters, i.e. those which have irregular shapes with narrow \emph{bridging regions} in multidimensional space. However, the main advantage is closely related to the main issue with the HCM, the so-called \emph{chaining} effect. That is, the first concentrations naturally tend to incorporate nearby groups, thus forming a \emph{chain}. In the case of nearby families, the chaining could prevent separating the families. The chaining effect is because only a single pair of points is needed to be close enough to merge two clusters.

\subsection{Removal of interlopers}
\label{ss:interlopers}

The HCM is a statistical technique that, along with real family members, unavoidable associates also some interlopers with a family \citep{1995Icar..118..271M}. The presence of these interlopers complicates any family-related study, and their removal is of great importance. Asteroid families are known to be mostly homogeneous in the composition \citep[e.g.][]{2008Icar..198..138P}, and therefore the members of a family share similar surface reflectance characteristics, including the spectra, albedos, and colours. If available, this information could be used to discriminate between actual family members and nearby background asteroids.\footnote{As a matter of completeness, let us mention that some interlopers could also be identified based on their position concerning the V-shape of the family \citep{2006Icar..182...92V,2015Icar..257..275S}. Such objects are located outside the family V-shape lines. Therefore, they are too far from the centre of the family to be transported to their present location by the Yarkovsky effect. This is illustrated in Section~\ref{ss:yarko_ages} for the Massalia family.}

The surface reflectance data could be used in different ways. One possible method to exploit these data is to apply the HCM in \emph{extended} space, i.e., in the space that includes surface reflectance properties along with the three proper elements. Following pioneering work by \citet[][]{nesvorny05}, several authors employed
this strategy. \citet[][]{2008Icar..198..138P} applied the HCM in four-dimensional space, using a linear combination of the Sloan Digital Sky Survey (SDSS) colours as the fourth dimension. \citet[][]{2013MNRAS.433.2075C} extend this approach, adding albedos as the fifth dimension, significantly reducing the percentage of known interlopers with respect to other methods. 

Another possible strategy is to separate the main-belt asteroids into two populations (typically representing C- and S-type objects) according to their colour or albedo values. The HCM is then run to each of these populations separately. \citet[][]{masiero13} utilize this technique onto the WISE albedos. They were able to link almost 40,000 asteroids to 76 dynamical families and identify several new families. 

A more recent example of this technique was applied by \citet{2017AJ....153..266N}.
These authors used available data to separate asteroids from the Phocaea region into C- and S-complex and then applied the HCM method to the sub-population of C-type objects. The approach allowed identification of the Tamara family, a group of dark asteroids embedded in the dominant population of bright and rocky S-type asteroids.

Both the above-described strategies are very useful but have limitations in that they can only be applied to a reduced set of main-belt asteroids for which the colours and albedos are obtained. Despite the significant increase of available physical data in recent years, the number of asteroids for which these data are at our disposal is still several times smaller than the number of objects for which proper elements have been computed. As \citet[][]{2014Icar..239...46M} explained, the proper elements contain more comprehensive information than the physical data because the latter are available either for significantly smaller catalogues or with lower relative accuracy. For this reason, some authors adopted a bit different approach, so that the so-called dynamical families are first obtained in the space of proper elements, and available physical data is used only posterior to identify interlopers among family members \citep[e.g.][]{2011Icar..216...69N, 2014Icar..239...46M}. While it allows identifying much more family members with respect to the previous approach, the pitfall of this strategy is that reflectance characteristics are of limited usefulness in separating the overlapping families.

A possible improvement in the interloper removal methodology from a different perspective was proposed by \citet{2017MNRAS.470..576R}. These authors developed a variation of the standard HCM approach, introducing an additional step in the procedure to reduce the above-mentioned chaining effect. The main idea is to prevent chaining through an already identified interloper. A family is first identified using the classical HCM approach in this two-step procedure. Then, potential interlopers identified based on their reflectance characteristics (e.g. spectrum, colour, albedo) are removed from the input catalogue, and the HCM is rerun. This prevents connecting new members to the family by linking them to a suspected interloper.
The obtained results show that the number of potential interlopers among family members could be significantly reduced in this way. However, a drawback is that this approach could miss a part of the family for families of a very complex shape.

Therefore, information about surface reflectance characteristics helps refine the membership of an asteroid family. Still, a sort of caution is needed here. Interloper removal methods based on the composition of asteroids associated with a family intrinsically assume a homogeneous family. Though available data indeed suggest that most families are homogeneous in composition, there are some exceptions to this. Such an approach that is apriori giving up the possibility of a heterogeneous family could prevent us from discovering heterogeneous families. It contradicts, in some way, claims that families provide us information about the parent's body composition. Braking this circular dependence is not trivial, but a determined fraction of interlopers might be a hint if we deal with a heterogeneous family originating from a partly differentiated body. Based on the statistical arguments, \citet{1995Icar..118..271M} found that the fraction of interlopers in an asteroid family identified using the HCM should be up to about $10\%$. Any fraction significantly larger than that could point out to a family heterogeneous in composition, or overlapping of two different families. This should be considered when a level of homogeneity/heterogeneity among family members is analysed.

\subsection{Recent advances in asteroid family identification}
\label{ss:identification_advances}

Dealing with big data sets has become the most challenging aspect of asteroid family identification in recent years. From a technical point of view, the classical HCM applied to a large number of asteroids becomes a very time-consuming procedure and, at the same time, requires a significant amount of available memory. Therefore, the problem is related to the computing resources needed to deal with many asteroids. Additionally, a large number of asteroids produce a high number-density of asteroids in the space of orbital elements causing many families to overlap. This is because the smaller family members are both launched at higher ejection velocities and faster transported away from the centre of the family due to the non-gravitational effects (see Section~\ref{s:dynevo}).

There are two main directions proposed so far to cope with the problem. A possible solution to deal with a large amount of data for the purpose of classification of asteroids into families could be a multi-step methodology proposed by \citet[][see also \citet{2016IAUS..318...28M}]{2014Icar..239...46M}. Alternatively, the problem could be treated by employing machine learning-based tools, which can be used either to identify new asteroid families \citep[][]{2019MNRAS.488.1377C,2020SerAJ.201...39V}, or to attach new members to known families \citep[][]{carruba20}.

Another point of view was brought into the game by \citet[][see also \citet{2017Sci...357.1026D}]{walsh13}. Instead of looking into the 3-D space of the proper orbital elements, the authors proposed identifying families using the so-called V-shape. The method is especially suitable for searching for very old and, therefore, highly dispersed families.

It also becomes increasingly important to find the best way to include different datasets available on asteroids. It refers to the development of the algorithms able to extract the maximum information from the available data, but also treat missing data as the different types of data are available for a different number of asteroids. A step forward in using an extended number of parameters to identify mutually related objects and treating missing data was recently proposed by \citet[][]{2021MNRAS.504.1571H}. The authors employed a cladistic approach to classifying the Jupiter Trojans. The main advantage of this method is that not all characteristics need to be known for an analysis to be carried out. This allows using more parameters without truncating the dataset due to missing values. 

In the following subsections, we describe the main properties of these new strategies and discuss their most important advantages and limitations.

\subsubsection{Multi-step approach}

A multi-step methodology in asteroid families identification is related to partitioning the asteroid data set into subsets typically based on asteroid sizes (absolute magnitudes). The general idea is to apply the HCM to larger objects first, and then include the smaller ones in subsequent steps. The most complete multi-step method was developed by \citet{2014Icar..239...46M}. Their procedure consists of the following main steps:

1. Divide the main belt into several zones to reduce the total number of objects which need to be treated simultaneously.

2. Remove objects having absolute magnitude $H$ fainter than a given threshold $H_{\rm limit}$. The limit typically corresponds to the observational completeness limit in a given zone, though in principle, other choices are possible as well.

3. Apply the HCM to each zone to the samples of asteroids with $H< H_{\rm limit}$. In this step, the so-called family \emph{cores} are identified.

4. Classify individual asteroids which had not been used in the core classification by attaching some of them to the established family cores. In this step, the asteroids having a distance from at least one core family member not larger than the critical distance are attributed to the core families.

5. Apply the HCM in each zone to the samples of \emph{intermediate background} asteroids, defined as the set of all the objects not attributed to any family in steps 3 and 4. 

6. Attach remaining background asteroids to all families using a single-step HCM. In this step, asteroids not associated with any family in steps 3 and 5 are attributed to all of these families if a distance from at least one family member is below the given critical distance. If an asteroid is attributed to more than one family, it is considered part of an intersection. 

7. Merging halo families with core families. There are essentially two possible cases for families identified at step 5: these families can either be fully independent, new families having no relation with the families identified previously, or they may be found to overlap with families from steps 3 and 4, forming \emph{haloes} of smaller objects surrounding some family cores. These two cases are distinguished based on information about intersection objects identified in step 6.

The above-described procedure is shown to be efficient in dealing with relatively large datasets. \citet[][]{2014Icar..239...46M} applied it to about 330,000 numbered asteroids, while \citet[][]{2016IAUS..318...28M} extended the analysis to multi-opposition objects yielding in total more than 500,000 asteroids. It also solved both problems, identification of new families and attribution of new asteroids to existing families. 

Still, in its present form, it would be very challenging to successfully apply a multi-step method to an order of magnitude more asteroids, which are expected to be discovered by ongoing and forthcoming sky-surveys over the next several years. A possible solution could be a further partition of the dataset and the introduction of new steps in the procedure. That would, however, probably make the process more difficult to follow and less reliable.

The main problem with this idea is that, in some cases, it splits members of an asteroid family into two (or even more) independent groups. Indeed, \citet[][]{2014Icar..239...46M} suggested a pathway towards clarification of such cases, but it could take years of monitoring to resolve such cases. Additionally, merging separate families based on the number of intersection asteroids is sometimes ambiguous and, at least in part, based on subjective judging.

\subsubsection{Machine learning-based methodology}

In the last decade, machine learning (ML) has emerged as one of the most promising complements to traditional data analysis and modelling methods in scientific fields. The first steps towards applying ML-based tools for identification of asteroid families and attribution of new members to existing families have been performed.\footnote{For a review on a more general application of machine learning to asteroid dynamics we refer the readers to review by \citet{2021arXiv211006611C}.}

\citet[][]{2019MNRAS.488.1377C} were the first to attempt to employ ML-based techniques to identify new asteroid families. The authors investigated the prospect of employing ML-based clustering algorithms available in the \emph{PYTHON} programming language for the identification of the asteroid families. In this work, \citet[][]{2019MNRAS.488.1377C} adopted the standard distance metric proposed by \citet[][]{1990AJ....100.2030Z} and focused on a sub-population of main-belt asteroids moving onto the high-inclination orbits. This preliminary investigation reveals the main advantages and disadvantages of utilizing ML tools for asteroid family identification. On one side, the algorithms are shown to be able to identify all significant groupings in the region and to provide results in shorter times than the traditional HCM approach. The ML-based HCM turns out to be also easier to use than the traditional one.

On the other hand, regarding the assessment of the membership of individual families, some limitations of the tested ML-based algorithms become apparent. The precision of the algorithm, defined as a ratio between the number of correctly identified family members and the total number of objects associated with a family, varies significantly from family to family. Though precision of 100\% has been achieved in some cases, it varies from $37$ to $100\%$, with a membership precision below 80\% for more than a half of families \citep[][Table 4]{2019MNRAS.488.1377C}.

Another point to consider is that \citet[][]{2019MNRAS.488.1377C} tested ML-based algorithms to identify families in less populous, high-inclination regions of the main belt. In this region, however, even the traditional HCM is still applicable due to the significantly smaller number of objects than the main belt's low-inclination part.

\citet[][]{2020SerAJ.201...39V} explored the possibility of using an artificial neural network (ANN) for the classification of asteroids into families. The obtained results showed that the ANN-based algorithm performs somewhat better than the clustering algorithms tested by \citet[][]{2019MNRAS.488.1377C}. In particular, the ANN was shown to be efficient in reproducing the traditional HCM results for large families with more than 1,000 members. The results for smaller families are found to be less reliable, suggesting that ANN may not consistently outperform the ML-based algorithms, and good practice would be to try both approaches whenever possible. An additional limitation of an ANN-based method is that it acts as a black box and that different runs could lead to different results that put the reproducibility of the results into question.

\citet[][]{carruba20} used ML methods to address another problem related to asteroid families, that is, the connection of newly discovered asteroids to existing families. This approach bears some similarities with the multi-step method. The basic idea is to identify families using only larger family members and then use the ML-based algorithm to attach new members to these families. Indeed, in principle, the second step could be applied to any family with known membership. Carruba and co-authors tested nine different ML-based classification algorithms. They found that the Extremely Randomized Trees method performs the best when attaching the new family members, typically achieving high precision between $83$ and $98\%$ \citep[][Table A6]{carruba20}. It shows that the ML-based algorithms could successfully attach new members to asteroid families. The algorithms perform better when attaching new members to existing families than when classification is done from scratch. We also note that supervised ML-based methods may overlook new patterns in the distribution of family members. These algorithms, trained on the orbital distributions of family members of known families, will look for similar patterns and therefore may not identify unusual structures.

\begin{figure}[h]%
\centering
\includegraphics[width=1.0\textwidth,angle=-90]{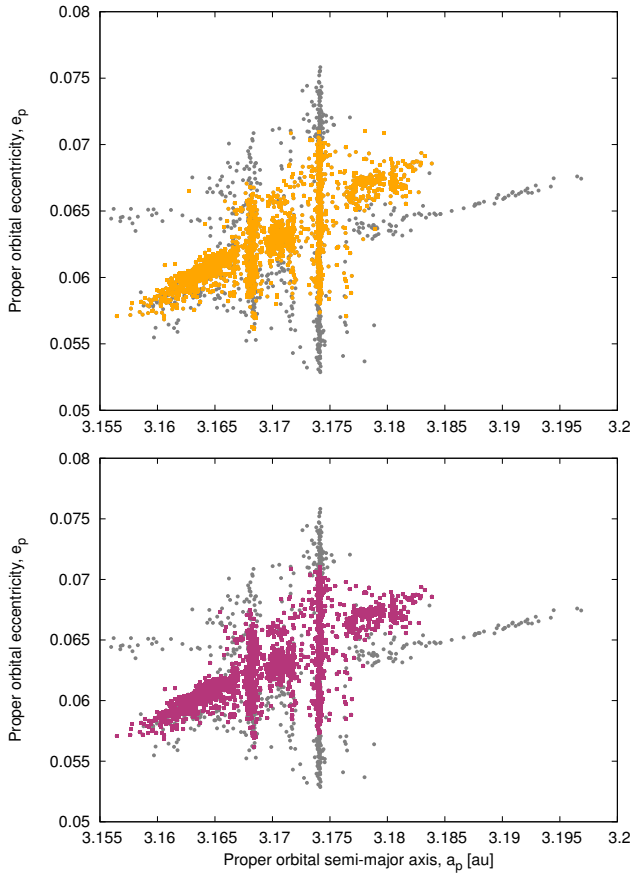}
\caption{Identification of Veritas family members using different methods and different catalogues. In both panels, grey circles represent the family members identified by the classical HCM approach applied to the catalogues containing $\sim$793k objects, respectively, and using the cut-off of 18~m~s$^{-1}$. 
Orange (upper panel) and magenta (lower panel) squares represent the members attached to the family from the more extensive catalogue, to the members initially identified with the classical HCM at 18~m~s$^{-1}$ applied to the smaller catalogue with $\sim$507k records. The upper panel (orange squares) are objects attached by the ExtraTrees ML-based algorithm with a hyper-parameter (\emph{number of estimators}) equal to 41 \citep{carruba20}. In the lower panel (magenta squares) are shown members attributed to the family with the single 18~m~s$^{-1}$ step HCM approach \citep{2014Icar..239...46M}.}\label{fig:ML_ext}
\end{figure}

In Fig.~\ref{fig:ML_ext} we show how the algorithms for attribution of new family members to an already existing family (ML-Based extra tree and one-step HCM algorithms) compare to each other, and how concerning the classical HCM approach. Applying the classical HCM at a cut-off of 18~m~s$^{-1}$ to an older (and therefore smaller) catalogue of proper elements that contains 507,449 objects, and to a larger catalogue with 793,310 asteroids, we identified 1452 and 2988 Veritas family members, respectively. In this case, applying the classical HCM to the catalogue that contains about $56\%$ more asteroids than the initial one, and keeping the cut-off distance at 18~m~s$^{-1}$, increased the family membership for 1536 members, i.e. by a factor of 2.

On the other hand, applied to the extended catalogue, the Extra Tree and one-step HCM algorithms respectively attributed 917 and 908 members to the family initially identified using the classical HCM and the smaller catalogue. From Fig.~\ref{fig:ML_ext}, it is apparent that the distribution of members attributed by the Extra Tree (orange squares in the upper panel) and one-step HCM (magenta squares in the lower panel) algorithms are almost identical. Therefore, these two algorithms show very similar performances.

The classical HCM algorithm attached a significantly larger number of objects when applied to the larger catalogue. However, the distribution of the identified members (grey circles in Fig.~\ref{fig:ML_ext}), shows very irregular patterns. As discussed in Section~\ref{s:identification}, this is at the same time related to the main advantage and the main disadvantage of the classical HCM. Though many families are definitely of irregular shapes, some of these structures might not be the real parts of the families, but rather interlopers associated with the family due to the chaining effect. The most severe problem that may arise due to chaining is merging two or more separate families.

Given the above example, both ML-base extra tree and one-step HCM algorithms perform pretty well in attaching new members to an existing family. Therefore, we recommend using one of these algorithms for fast checking of new family members. The algorithms could be especially useful in the areas of a very high number-density of asteroids. On the other hand, the classical HCM approach could still overperform the two attribution algorithms. It still seems like the best choice in cases where the number-density of asteroids and distances between nearby families allow its application.

\subsubsection{Primordial families and V-shape based detection of families}

The classical HCM approach typically fails to identify very old families, formed more than about 3 Gyr ago. Such ancient groups are so dispersed in the space of proper orbital elements that the HCM cannot distinguish them from the background population of asteroids. For this reason, many current efforts have focused on detecting ancient families using different techniques. 

Recent methods make use of the V-shape (see Section~\ref{ss:yarko_ages}) to identify families that are too dispersed to be discovered using the classical HCM. The idea is to test for possible V-shape number-densities amongst the background of unassociated asteroids. V-shape finding algorithms can
be divided into two categories. The first is the so-called \emph{border method} (aka $dC$-method or $dK$-method) that utilises a ratio between the number of objects just outside and inside the
V-shape. It is helpful when families have a distinct edge and stand out with respect to the background asteroids
\citep[see][]{walsh13,2017Sci...357.1026D}.  The second category refers to the
\emph{density methods} (aka $a_w$-method). The goal is to detect a peak of asteroid density
in a V-shape plane. It looks like a better technique for families
embedded in a dense background of asteroids~\citep{bolin17}.

These V-shape based family identification techniques are not trivial to implement.
Here we review the main steps, while for full implementation details, we refer readers to
\citet{bolin17} for the implementation of the $dC$, and \citet{2017Sci...357.1026D} for the $a_w$ version of the method.

The $dC$-method or $dK$-method is part of the border method
category. It is based on testing a family centre and age or slope (how
large the V-shape is). This technique tests a set of family centres and
family slopes, evaluating a ratio between the bodies just inside and outside the defined V-shape borders. They are called $K$
or $C$ methods because the slope $K$ relates to the parameter $C$
through $K=\sqrt{p_v}/(D_0C)$, where $p_v$ is the geometric visible
albedo, $D_0 = 1329$ km~\citep{walsh13, 2017Sci...357.1026D}, and $C$ is the
parameter defining the re-scaling asteroid's distance from each
other~\citep{vetal2006}. The $a_w$-method considers a slightly different approach. It scans for
the V-shape vertically in the $(a_{\rm p},1/D)$ or $(a_{\rm p},H)$ space and calculates
the ratio of asteroids above and below the tested V-shape. The offset
is established as a delta in $1/D$ and named as $a_w$~\citep{2017Sci...357.1026D}.
\citet{deienno21} summarizes the two techniques saying that the
$dC/dK$-method searches for $(K \pm \Delta K) \lvert a_{\rm p}-a_{{\rm p}, c} \rvert$ and the
$a_w$-method searches for $(K \lvert a_{\rm p}-a_{{\rm p}, c} \rvert \pm a_w)$.

Both categories have been tested on synthetic
families~\citep{walsh13,bolin17,deienno21} and they have also been
used to try to detect primordial families. \citet{walsh13} were the first to use
the V-shape based family identification finding the \rm{new Polana family}, about $2$~Gyr old group, and the Eulalia family, about $0.8-1$~Gyr old group, both within
the Nysa-Polana complex. ~\citet{2017Sci...357.1026D} proposed 
a $4$-billion years old family extending across the entire inner part of
the main belt, whose members include most of the dark asteroids
previously unlinked to families.~\citet{delbo19} extended the research
of ancient families, suggesting other two ancient families, Athor and
Zita. 

A similar technique was used by \citet[][]{2015Icar..252..199D}, extending the work done by \citet{walsh13}, to study the Nysa-Polana complex of families. In addition to seeking clusters in the dynamical and reflectance parameter space,
the authors seek signatures of dynamical evolution that distinguish multiple families within a single group. \citet[][]{2015Icar..252..199D} have identified five distinct clusters of asteroids within the Nysa-Polana orbital region of the main belt. Two of these clusters are associated with collisions on (135) Hertha, two others with (495) Eulalia, and one is linked with (142) Polana.

These different techniques allow identifying some very old families in the inner main asteroid belt. However, the V-shape family identification method is also very sensitive to many parameters involved in the search, and additional verifications are needed to fully quantify the robustness of the method.

\citet{deienno21} have done a comprehensive study to characterise
the efficiency of the V-shape method in detecting asteroid families. They
worked using a synthetic population of asteroid families and background
asteroids, letting those families evolve over billions of years, setting different levels of ratio of the synthetic family and background asteroids to derive a detection efficiency map for the V-shape method. They also tested their methods using the families found by~\citet{2017Sci...357.1026D,delbo19}. The paper results show that families older than $3$ Gyr are difficult to detect, while families $0.5$-$2.5$ Gyr old are more easily detected, with an efficiency of more than $80\%$. In light of these results, \citet{2017Sci...357.1026D,delbo19} seem to be lucky to detect the ancient families. As stated by \citet{deienno21}, this is still a work in progress, even though promising. Other steps will include adding statistical analysis and other debiasing techniques to fully characterise the V-shape methods' efficiency. Especially, the methods need to be tested for their reliability, that is, how often they could potentially generate false-positive results. 

\subsubsection{Cladistic approach}

A step forward in using an extended number of parameters to identify mutually related objects and treating missing data was recently proposed by \citet[][]{2021MNRAS.504.1571H}. The authors employed a cladistic approach to classifying the Jupiter Trojans.

The main advantage of this method is that not all characteristics need to be known for a cladistical analysis to be carried out. This allows using more parameters without truncating the dataset due to missing values. 

The method, however, has some setbacks as well. First, the algorithm is time-consuming and scales badly with the number of objects. Therefore, its extension to the main belt region would be very challenging. Second, though the ability to treat the missing data is one of the method's main advantages, it should be noted that such asymmetric datasets could lead to spurious or unreliable results in some cases. For instance, if only basic information is available, such as apparent magnitudes and orbit solutions, the amount of data is insufficient to place the significant weight on the results. Nevertheless, the cladistic approach employed by \citet[][]{2021MNRAS.504.1571H} represents an important step forward in analysing mutual relationships among the objects in an asteroid population. In order to take full advantage of the approach, caution is needed when interpreting the data. A possible way to increase the robustness of the results could be achieved by combining cladistic techniques with other available methods.

\subsection{Asteroid Families Portal}
\label{ss:afp}

For the reasons explained above, identifying new families and the association of members to existing families are demanding procedures that require careful analysis, and the results of such studies are typically published as journal papers. However, a fast-growing number of newly discovered asteroids facilitates also need to constantly update the list of know families, or at least to associate new members to existing families. In recent years, web portals and databases have been recognized as a possible way to cope with the large size and complexity of data accumulated over the years. Such portals are often devoted to specific purposes and allow scientists to quickly access and review different data. With the possibility of integrating the data and the computation, such portals enable interactive analytics, typically based on well-established tools and algorithms. They, on one side, allow better exploitation of the available data, while on the other side, make using different tools simpler and available to a broader community.

In 2017, an internet portal on asteroid families was launched, namely the Asteroid Families Portal (AFP). The portal is freely available at \url{http://asteroids.matf.bg.ac.rs/fam/}, and provides data and tools to study asteroid families, including some tools described in Section~\ref{s:identification}. The primary goal of the AFP is to implement various algorithms used to study asteroid families and, therefore, to allow users to employ these tools online. Additionally, the portal collects different data about families and closely related subjects.

\begin{figure}[h]%
\centering
\includegraphics[width=0.8\textwidth,angle=0]{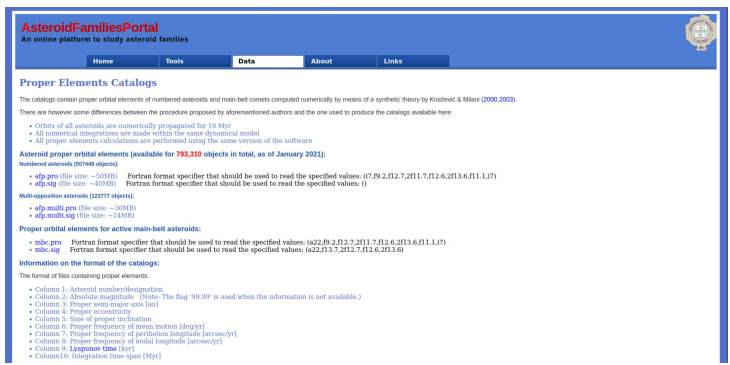}
\caption{Snapshot of the \emph{Proper Elements} page at the Asteroid Families Portal. The page is available at: \url{http://asteroids.matf.bg.ac.rs/fam/properelements.php}}\label{fig:afp_pro_screen}
\end{figure}

Currently available services and data at the AFP covers the following five categories:
\begin{itemize}
\item Application of the Hierarchical Clustering Method to identify members of an asteroid family: as explained in Section~\ref{s:identification}, the HCM is the most widely used method for the identification of asteroid families and their memberships. The AFP allows the algorithm described in \citet{1990AJ....100.2030Z} to be run online using either default parameters (for less experienced users) or advanced settings (for more experienced users).
\item Application of an automatic procedure to obtain a list of family interlopers: this option is based on the algorithm described in \citep[][]{2017MNRAS.470..576R}.
It provides the list of potential interlopers and the information used to suggest such status of an asteroid.
\item Application of the Backward Integration Method to estimate the age of an asteroid family: for a given list of asteroids, the algorithm described by \citet{netal2003} is run over the time series of mean orbital elements. It returns, as a result, the time evolution of average differences in secular angles (longitude of node $\Omega$ and longitude of perihelion $\varpi$) for selected objects. The analysis covers an interval of 10~Myr in the past.
\item Proper orbital elements of numbered and multi-opposition asteroids, and active asteroids: the synthetic proper elements available at the AFP are computed numerically following, in principle, the method described in \citet{2003A&A...403.1165K}. The important difference with respect to the procedure used by \citeauthor{2003A&A...403.1165K}, is that orbits of all asteroids are numerically propagated for 10 Myr, and using the same dynamical model that includes seven major planets (excluding Mercury). Fig.~\ref{fig:afp_pro_screen} shows the snapshot of the page at the AFP devoted to the proper elements.
\item List of the peer-reviewed papers on asteroid families: the currently available data goes back to 2007, intending to extend it more in the past.
\end{itemize}

\subsection{Future prospects and challenges for asteroid families identification}

Identifying families and classifying individual asteroids into families is already a very challenging task. In a way, paradoxically, these challenges are mainly due to a large amount of newly available data, especially newly discovered asteroids. The problems will soon become even additionally severe once more extensive data sets produced by the large sky-surveys become available, especially when the Vera Rubin telescope will start observing~\citep{ivezic19_lsst,lynn21_lsst}, discovering millions of new small solar system objects.

On a purely computational side, the classical HCM method is not suitable for dealing with many objects. It is primarily not due to long computation time but rather because of extreme memory requirements. These limitations are challenging, but there are opportunities to overcome them, especially the execution times, by newly available clustering techniques. Even already developed methods presented in Section~\ref{ss:identification_advances} could overcome these issues to some degree.

A more difficult problem is a large number density of asteroids, which reduces the average distance between two objects below the uncertainty in the distance computation. A large number of objects will also further increase the chaining effect, which will cause families to overlap in the space of orbital elements. These problems will be tricky to solve. 

\begin{figure}[h]%
\centering
\includegraphics[width=\textwidth,angle=0]{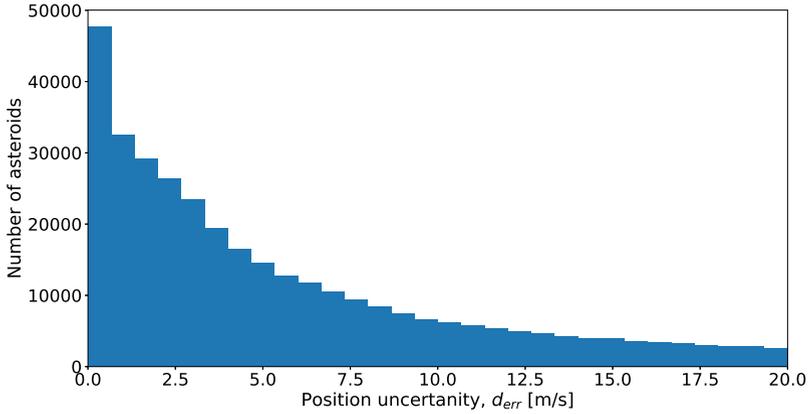}
\caption{Position uncertainties distribution in the space of the proper orbital elements, expressed in meters per second. The data for about 507,000 numbered asteroids are shown. The errors are computed according to Eq.~(\ref{eq:hcm_dis}), using the formal uncertainties in proper semi-major axis, eccentricity and sine of inclination.}\label{fig:pro_err}
\end{figure}

Despite being typically of much better relative accuracy than the physical parameters on asteroids, uncertainties in the proper elements further complicate the situation. In Fig.~\ref{fig:pro_err}, we show the distribution of the position uncertainties caused by uncertainties in proper orbital elements. The uncertainty in the position of an individual asteroid is computed using the standard HCM metrics given by Eq.~(\ref{eq:hcm_dis}). These uncertainties in the position of individual objects translate into uncertainties in the distances between the asteroids. If significant, such uncertainties further complicate classifying asteroids into families. Until recently, these uncertainties were safely below the typical distances among the asteroids in the 3D space of proper elements. However, a recent increase in the number of known asteroids significantly decreased the average distance between the two neighbouring objects, making position uncertainty an important issue. It needs to be taken into account in future works related to identifying asteroid families and attributing new members to known groups.

Another aspect of the family identification problem is the generally unknown distribution of the background (non-family) objects. A standard assumption behind applying clustering algorithms to identify asteroid families is the uniform distribution of the background (non-family) objects. It is well known, however, that this is not the case. There are at least two reasons for that. First, dynamical evolution removes faster objects from some regions of the main asteroid belt than from others. Second, the procedures used to compute the synthetic proper elements introduce some non-uniformity by themselves. Good examples are concentrations of asteroids near the centres of the resonances, which is a consequence of an averaging procedure used to compute the proper elements. A potential way to reduce the effect of this issue is to use a synthetic background that better represents a \emph{random} population in the proper elements space.

\begin{figure}[h]%
\centering
\includegraphics[width=0.8\textwidth,angle=0]{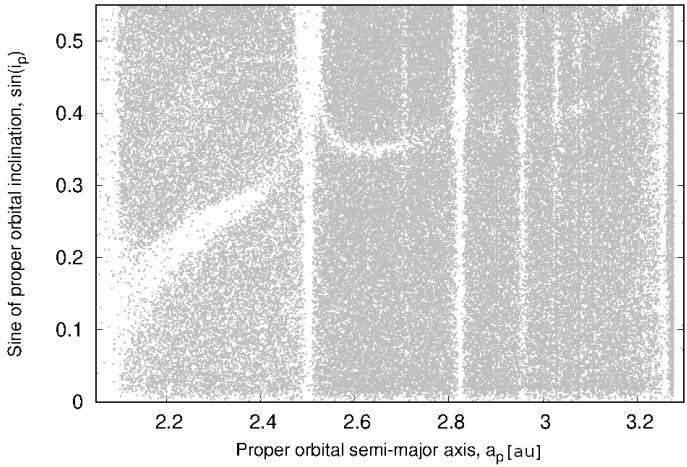}
\caption{Proper elements distribution for about 200,000 test particles uniformly distributed across the main belt in the osculating orbital elements.}\label{fig:fic_pro}
\end{figure}

Such a population could be constructed starting from a uniform distribution in osculating elements space and computing the corresponding proper elements for a number of test particles. The outcome of such an approach is shown in Fig.~\ref{fig:fic_pro}.

The most difficult to solve is probably the problem of the overlapping families. To deal with this issue, we need to look in another direction. For instance, identifying families in multi-dimensional space (N$>$3) could be an option, but appropriate variables and metrics need to be defined. For instance, combining the standard metrics with the V-shape criterion could help a bit. However, this would apply only to families older than about 50~Myr. Also, for one-side families, it may not work well. On the other hand, in the case of overlapping families, the main problem is not necessary to recognize that there are more than a single family, but the issue is how to attribute individual asteroids to these groups correctly. These remain open challenges for future works.

Certainly, identifying asteroid families and attribution of new members to existing families will be an important task in the near future. We discussed here the main problems and proposed some of the directions towards their solution. However, more comprehensive work is needed to understand better which techniques and algorithms will or will not work.

\section{Long-term dynamical evolution}
\label{s:dynevo}

To successfully reconstruct the long-term dynamical evolution of an asteroid family, it is essential to understand all the relevant mechanisms that change the family and introduce these effects into a model. The relevant dynamical mechanisms could be broadly divided into the gravitational and non-gravitational effects. To account for the gravitation effects, generally, the best strategy would be to include the gravitational influences of all the solar system objects in the model. Due to computational reasons, this is, however, not feasible. Therefore, a balance between the approach needs to be followed. Luckily, for the main-belt asteroids, gravitational perturbations of most of the solar system objects are negligible. The relevant perturbations are typically caused by eight major planets and a few of the most massive asteroids. In some cases, even not all these objects need to be taken into account. Regarding the non-gravitational effects, for asteroids, relevant are the Yarkovsky and Yarkovsky-O'Keefe-Radzievskii-Paddack (YORP) effects. Except due to the dynamical evolution, families evolve for other reasons, and collisional evolution is especially important. The dynamical and collisional evolutions are coupled, and simulations of the dynamical evolution often need to account also for the collisional evolution.

The simulations of the long-term evolution of asteroid families have seen many remarkable achievements over the years. There is a long list of literature covering these topics. It is beyond our goal to mention all these works. Instead, we primarily focus on the results published since 2015 and briefly describe some of the key results from the more distant past. A more in-depth review of the results achieved in the more distant past can be found elsewhere \citep[see for instance][and references therein]{2009P&SS...57..173C,2015aste.book..297N}. For a review of the dynamics of families interacting with secular resonances, we refer the readers to the recent review by \citet[][see also \citet{2022SerAJ}]{2018P&SS..157...72C}.

Probably the most important step forward in modeling the long-term orbit evolution of asteroid families, made in the last several years, was an introduction of the \emph{stochastic YORP} concept. The concept, based on the work by \citet[][]{2009Icar..202..502S}, was introduced by \citet{2015Icar..247..191B} in their search for a possible origin of asteroid Bennu. The authors used a Monte Carlo approach to study the dynamical spreading in terms of the orbital semi-major axis of five inner main-belt families. \citet{2015Icar..247..191B}, however, assumed that due to shape changes to asteroids, produced by processes such as crater formation or changes to asteroid rotational angular momentum by YORP, may cause the asteroids' spin rates to undergo a random walk. The \emph{stochastic YORP} mechanism slows down how often asteroids reach YORP end-states. The effect, in turn, allows the faster drift of the asteroids' semi-major axis than in the classical Yarkovsky/YORP scenario. This new model successfully reproduced the semi-major axis distribution of the observed families, and opened possibilities to explain some other mismatching related to the modeling of the YORP effect.

Among the recent highlights was also the discovery of the dynamical importance of secular resonances with massive bodies other than the major planets. \citet{2015ApJ...807L...5N} found that the spreading in orbital inclination seen in the Hoffmeister family is an outcome of the $\nu_{1C} = s - s_C$ nodal linear secular resonance with Ceres. The authors showed that passing through the $\nu_{1C}$ resonance may cause significant changes in the orbital inclination of an asteroid. It was the first direct proof that a secular resonance between Ceres and other asteroids, previously completely overlooked, can cause significant orbital evolution. 

\begin{figure*}[]
\centering
    \includegraphics[width=0.97\textwidth,angle=-90]{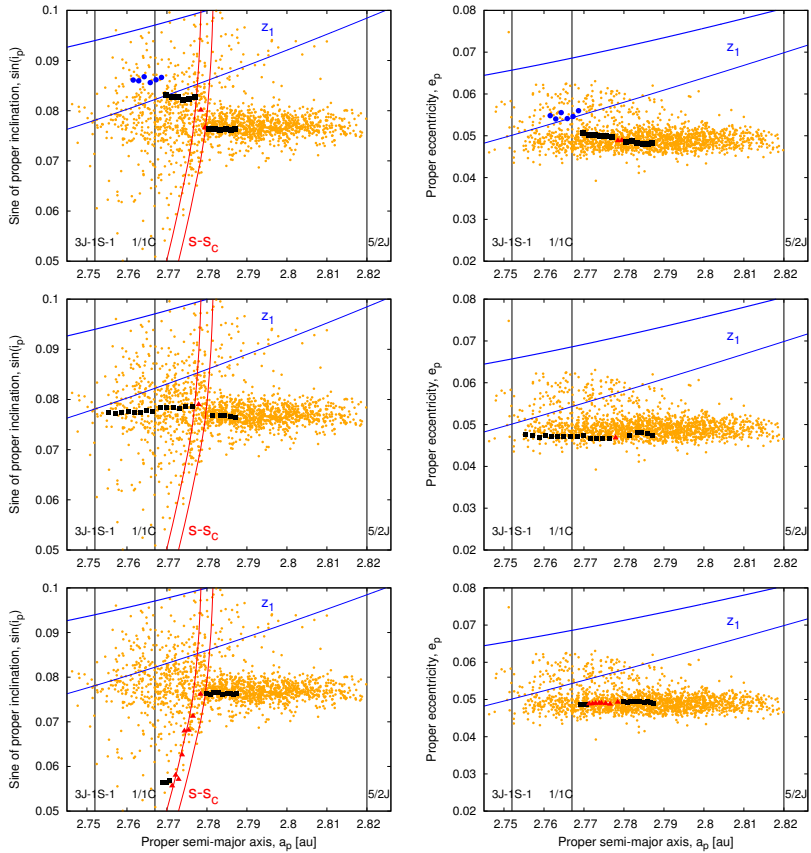}
    \caption{Three representative scenarios (top, middle and bottom panels) of the orbit evolution of an asteroid from the Hoffmeister family. The left and right panels show the evolution in the semi-major axis vs inclination ($a_{\rm p}$, $\sin(i_{\rm p})$) and the semi-major axis vs eccentricity $a_{\rm p}$, $e_{\rm p}$)  planes, respectively. The examples highlight the interaction of the Yarkovsky effect and the $\nu_{1C}$ secular resonance with Ceres. The dashed dark red line represents the approximate location of $\nu_{1C}$ resonance. The full and dashed blue lines represent the centre and one border of $z_{1}$ secular resonance, respectively. Thin vertical dashed lines indicate the mean motion resonances. Grey dots represent the distribution of the family members after 200~Myr evolution. Orange squares show the successive locations every 10~Myr of a selected test particle. When the $\nu_{1C}$ catches the particle, the orange squares change into dark red triangles, while inside the $z_{1}$, the particle is represented by blue circles.}\label{f:scenarios}
\end{figure*}

To illustrate the principle of the dynamic evolution simulation and better explain the role of secular resonances with Ceres, we discuss the latter in more detail. For this purpose, some possible scenarios of the dynamic evolution of the Hoffmeister family members are shown in Fig.~\ref{f:scenarios}. The figure shows the orbit evolution of test particles over 200~Myr. Three representative test particles are selected with negative Yarkovsky drift and, therefore, drifting toward the Sun. They are selected to show the interaction with the $\nu_{1C}$ secular resonance. In the top panels, we see an example of the particle drifting only in terms of the proper semi-major axis for about 65 Myr, when it reaches the $\nu_{1C}$ resonance. The interaction with the resonance increased the orbital inclination of the test particle. Once the particle exits $\nu_{1C}$, the Yarkovsky effect drives it to $z_{1}$ in about 60 Myr. After that, $z_{1}$ offers a path followed by the asteroid until the end of the simulation.

The middle panels show the evolution of an asteroid with the fastest negative Yarkovsky drift used in the simulations. In this case, the particle went across the resonance in a relatively short time, about 10~Myr, and interaction with the $\nu_{1C}$ only slightly increased the orbital inclination. Once outside the resonance, the particle continues drifting in the semi-major axis. The orbital eccentricity and inclination are not affected, suggesting that the particle is not interacting with the $z_{1}$ resonance.

The bottom panels illustrate the evolution of an asteroid whose proper inclination is significantly reduced by the $\nu_{1C}$ resonance. It reduces approximately 0.021 in the sine of proper inclination, meaning a loss of 1.2 degrees in proper inclination in less than 100~Myr. Such a scenario is favoured by the orientation of the $\nu_{1C}$, which creates a path in agreement with the inward direction of the Yarkovsky drift. While the inclination is dropping down, eccentricity remains approximately constant. This behaviour is in agreement with the fact that the particle remains far enough from $z_{1}$, so that this resonance cannot influence the motion of the particle.

The work by \citet{2015ApJ...807L...5N} opened several new possibilities to investigate small bodies' dynamics and initiated related studies. An interesting aspect in this respect is that by modifying the orbital inclination of family members, the nodal secular resonances indirectly modify the initial ejection velocity field, particularly an out of the orbital plane velocity component. As a result, the distribution of this velocity component for members of an asteroid family becomes more and more leptokurtic\footnote{A leptokurtic distribution has positive kurtosis larger than that of a corresponding Gaussian distribution. In general, leptokurtic distributions have heavier tails or a higher probability of extreme outliers values when compared to Gaussian distributions.} over time \citep[][]{2016MNRAS.463..705C}. This fact could be, in turn, used to estimate the initial (post-impact) velocity distribution. For details on the studies along these lines, see Section~\ref{ss:ini_vel_field}. 

The Astrid family members have an unusual orbital inclination distribution, the feature caused by an interaction of the family with the $\nu_{1C}$ nodal secular resonance with Ceres \citep[][]{2016IAUS..318...46N}. Consequently, the observed distribution of the $v_W$ component of terminal ejection velocities is quite leptokurtic. \citet{2016MNRAS.461.1605C} used the orbital configuration of the Astrid family to set constraints on the parameters describing the strength of the Yarkovsky force, such as the bulk and surface density and the thermal conductivity of surface material. By varying the initial conditions, and by demanding that the current value of the kurtosis of the distribution in $v_W$ be reached over the estimated lifetime of the family, \citet{2016MNRAS.461.1605C} found the thermal conductivity of the Astrid family members should be $0.001$~W~m$^{-1}$~K$^{-1}$, and that the surface and bulk density should be higher than 1000 kg~m$^{-3}$. Furthermore, using the Monte Carlo approach to simulate the Yarkovsky/YORP evolution of the family, the author estimated its age to be $140\pm30$ Myr. Similar arguments were used by \citep[][]{2017MNRAS.465.4099C} to reassess the Hofmeister family. By combining the results from a Yarkovsky/YORP Monte Carlo method with the constraints provided by the inclination distribution and the time behaviour of the kurtosis of the $v_W$ component of the ejection velocity field, the authors found that the Hoffmeister family should be 220$^{+60}_{-40}$~Myr old.

\citet{2019P&SS..165...10C} studied the dynamics of the Nemesis asteroid family, also considering the influence of Ceres. An especially unusual characteristic of the family is the asymmetric distribution of its members in terms of the orbital semi-major axis. More than $85\%$ of the members are located at smaller semi-major axes with respect to the centre of the family. The authors, however, found that the asymmetric shape of this family is probably an outcome of the original ejection velocity field, rather than the post-impact dynamical evolution. Considering the population of family members currently present in the $\nu_{1C}$ resonance, \citet{2019P&SS..165...10C} set an upper age of the Nemesis family to be 200~Myr.

Regarding the interaction with $\nu_{1C}$ secular resonance, an interesting case for further study is the Seinajoki asteroid family. The family is also crossed by this resonance \citep{2016Icar..280..300T}, and its overall shape in the $(a_{\rm p}, \sin i_{\rm p})$ plane could be explained by the effects of the $\nu_{1C}$ resonance \citep[][]{2016IAUS..318...46N}. However, although the perturbation caused by this resonance explains the overall shape of the Seinajoki, several smaller features remain unexplained. \citet{2015Icar..257..275S} found that the V-shape boundaries of the family are consistent with two different ages, suggesting that the family might be an outcome of two collision events.

The proper orbital elements may be especially suitable for studies of the long-term dynamical evolution of asteroid families. These are the quasi-integrals of the full N-body equations of motions and, as such, nearly constant in time, thus preserving the dynamical signatures from the remote past \citep[see e.g.][and references therein]{2017SerAJ.195....1K}. For these reasons, the proper elements (or their approximation as long-term averages) are often used to follow the dynamical evolution of asteroid families. 

\citet[][]{2002Icar..157..155N} were the first to study the long-term evolution of an asteroid family in the space of proper elements. The authors used averaged elements over a long time as a proxy for the proper orbital elements. The long-term evolution of the Flora family was simulated by combining shorter N-body simulations with a kind of Monte Carlo simulation. In this paper, Nesvorn{\'y} et~al. focused on the dynamical dispersion of the proper eccentricity and inclination, which occurs due to the chaotic diffusion in narrow mean motion resonances, and considered the Yarkovsky non-gravitational force and the gravitational impulses received at close approaches with large asteroids. More recently, \citet[][]{2010CeMDA.107...35N} employed a similar Monte Carlo approach to study the evolution of the Lixiaohua family in the space of proper elements. The authors used a model which combines the chaotic diffusion, Yarkovsky thermal effect, and random walk due to close encounters with nearby massive asteroids. These simulations showed that all the effects should be taken into account to reproduce the observed distribution of family members accurately.

Still, although the two above analyses were performed mimicking the evolution in the space of proper elements, these were based on Monte Carlo techniques. The evolution of asteroid families in the space of proper elements but based on the full N-body simulations were performed by \citet[][]{2015ApJ...807L...5N} to study the Hoffmeister family, and more recently by \citet[][]{2017AJ....153..266N} to study the Tamara family.

Recently, \citet[][]{2019Icar..317..434B} developed an improved method to study the 3-dimensional shapes of asteroid families in the space of the proper elements. The method relies upon comparing the observed and synthetic families from N-body simulations as directly as possible. A primal novelty of the approach is in simultaneous modelling of the evolution of both a family and background population of asteroids. \citet[][]{2019Icar..317..434B} applied their model to the Eos family and were able to explain most of the family's features fully. 

The long-term evolution of asteroid families due to non-gravitational effects may produce specific patterns in family members' distribution. The well-known phenomenon is the formation of family ears \citep[][]{2006Icar..183..349V,2020AJ....160..127L}. The ears form because the small family members concentrate toward the extreme values of the semi-major axis. At the same time, there is a lack of asteroids in the family center, giving the family the appearance of ears. This feature is a consequence of the YORP effect, which produces a perpendicular spin axis orientation relative to the orbital plane and maximizes the Yarkovsky drift. This expected central depletion is, however, not always clearly noticeable. \citet[][]{2016Icar..274..314P} investigated this feature in detail and developed a concept of the \emph{YORP-eye} and a central depletion parameter. The analysis allowed the authors to find YORP footprints in several families. Furthermore, \citet[][]{2016Icar..274..314P} also found that the location of the YORP-eye within the family evolves with a family. Therefore, the \emph{YORP-eye} could be used as an indication of the age of the family (see Section~\ref{ss:yarko_ages} for further details on the age estimation methodologies). 

An excellent example of what we can currently achieve in modelling the dynamical evolution of an asteroid family is a recent study of the Clarissa family by \citet[][]{2020AJ....160..127L}. In this work, the authors primarily aimed to apply the Yarkovsky effect chronology to estimate the age of the family. In other words, the goal was to define the appropriate initial conditions and follow the dynamical evolution in time until the observed orbital structure of the Clarissa family is matched. The unique challenge, in this case, was that it is highly asymmetrical in the proper semi-major axis. Therefore, to successfully reproduce the shape of the family, \citeauthor{2020AJ....160..127L} ran multiple simulations by varying initial ejection velocity fields and the fraction of initially retrograde spins. 

Additionally, the authors tested different Yarkovsky/YORP models. In particular, different YORP models were used by changing how often the effect tends to accelerate or slow down the rotation. Interestingly, the best fits were obtained with the model assuming a 4:1 preference for spin-up by YORP, and assuming that about 80\% of small family members initially had retrograde rotation. The corresponding age of the Clarissa family was found to be $56\pm6$ Myr. The work by \citet[][]{2020AJ....160..127L} demonstrated how a relatively complex situation could be solved with careful modelling and extensive testing.
\smallskip

\noindent{\it Limitations of the current models.-- }One of the most critical challenges that we face in modelling the long-term dynamical evolution of families is still the proper modelling of the YORP effect. It changes the spin axis orientation and spin rate of asteroids and therefore affects the Yarkovsky effect because the magnitude of the latter depends on the rotation state of asteroids. However, the YORP effect depends on asteroid shapes and small scales topography, such as the presence of boulders and craters at the surface. This already complex situation is further complicated by the fact that non-destructive collisions could also change spin states. It implies that to model the Yarkovsky effect adequately, it is necessary to model the combined effect of the YORP and impacts. On one side, we are dealing with impacts that are stochastic events and could be treated only in a statistical sense. On the other hand, the YORP effect could be decomposed into a stochastic and deterministic component. The modelling of the deterministic component is improving over the years, and new models are developed \citep[see][and references therein]{vokrouhlicky-etal_2015,2019AJ....157..105G,2021AJ....162....8G}. On the other hand, the contribution of boulders, craters, and shape evolution to the YORP effect is generally not deterministic. The distributions of craters and boulders, and their evolution at the surface of an asteroid are generally unknown. 

If we only consider the deterministic component of the YORP effect, the spin axis evolution due to the YORP typically reduces the magnitude of the Yarkovsky effect. It is because, for about a half of asteroids, YORP should bring obliquities to 90 degrees \citep{2021AJ....162....8G}; see, however, also older results of \citet{cv2004} which predict the asymptotic YORP obliquities near extreme values of 0 and 180 degrees. The pole orientation near the orbital plane switches off the diurnal component of the Yarkovsky effect, which scales with $\cos \gamma$, with $\gamma$ being an asteroid obliquity. For another half of objects, the $\gamma$ is pushed towards 0 or 180 degrees what should, in principle, increase the Yarkovsky effect. However, the obliquities are not expected to stay at these asymptotic values, but instead, undergo continuous cycles over relatively short timescales. As a result, according to this theory, for most asteroids, YORP should reduce the Yarkovsky effect over a time scales significantly longer than the YORP cycle timescale.   

Such a scenario, however, does not match the observations. Though not related to the families, an excellent example of this is the influx of near-Earth asteroids, which is far too low to match the observations when the YORP effect is taken into account \citep[][]{granvik-etal_2017}. Therefore, the stochastic contribution of the YORP effect and spin evolution due to the impacts need to be added to the model. This, however, is not so simple, and authors are using different models that yield different results and sometimes even opposite conclusions.

For instance, despite its success in modelling some asteroid families and providing a reasonably good match to the observations, the \emph{stochastic YORP} model by \citet[][]{2015Icar..247..191B} involves some free parameters which are specially designed to justify the model to the observations. These are $c_{\rm YORP}$ which deﬁnes the strength of the YORP effect, and $c_{\rm reorient}$ that is an empiric multiplication factor that controls the characteristic timescale of reorientation events.\footnote{See \citet{2022arXiv220213656F} for a recent implementation of these effects into OrbFit and Mercury numerical integration packages.} Therefore, the approach is still missing a complete physical interpretation.

Recently, \citet[][]{2020AJ....160..128M} presented in some sense the complete model of the evolution of asteroid families under Yarkovsky/YORP effects and collisions. Apart from being the most complete so far, the model has the advantage of not involving nonphysical free parameters. However, despite being promising at first, the model is yet to be verified against the observational constraints, as the authors tested their model only using the synthetic families. However, some of the presented results suggest that the observed spreading of the synthetic families could be far too small to match the spreading of real families, except assuming much older ages of these families. For instance, the maximum spreading of test particles assumed to be about 5~km in diameter, after $500$~Myr of a Koronis-like synthetic family evolution is only about $0.02$~au at the inner and even less at the outer side of the family \citep[Fig.10;][]{2020AJ....160..128M}. This translates into a maximum drift of $2 \times 10^{-4}$ au~Myr$^{-1}$, for a $1$~km size body. It is significantly below the theoretically predicted maximum Yarkovsky induced drifts or those derived, for instance, by scaling from the value obtained for asteroid Bennu, which suggests a maximum drift faster than $3\times 10^{-4}$ au~Myr$^{-1}$. Indeed, \citet[][]{2020AJ....160..128M} used the combined Yarkovsky/YORP model, which is expected to produce a somewhat smaller semi-major axis drift than the purely Yarkovsky effect. However, the level of the slowdown seems not to be supported by observational evidence.

Also, \citeauthor{2020AJ....160..128M}  did not observe any formation of the \emph{YORP-eye}. They suggested that this feature could be related to collisional physics rather than to the subsequent evolution driven by the Yarkovsky effect. However, this fact could also point out the model's limitations.

Despite all the above-described efforts and remarkable results in reconstructing the past evolution of asteroid families and their members, there are still many open questions and models to be improved. At the same time, it is essential to realize that asteroids have been subject to a large number of diverse stochastic events. Therefore, subsequent collisional and dynamical processes have masked information about the fragments produced by past collisions. It means the initial conditions and the post-impact evolution for many asteroid families (and especially the old ones) may never be precisely known. Given these limitations, modellers do the best they can with available information. Choosing parameters and formalism within the bounds of what is known and testing the results against the available constraints is a recipe to follow. Though, the interpretation of even good matches must always be met with some scepticism and wariness. Besides, a careful modeller needs also to run simulations over numerous trials to characterize how random events may have influenced outcomes. 

\subsection{Constraints on the initial ejection velocity fields}
\label{ss:ini_vel_field}

Reconstruction of an initial ejection velocity field (EVF) of asteroid families is essential to better understand the collisional events' physics. Such information could be cross-validated with the outputs from impact hydrocodes simulations \citep[][]{michel15_asteroidIV}. In turn, this may help, for instance, improve the asteroid internal structure models. However, evaluating an initial EVF of an asteroid family is a complex issue. Having said that, let us note that the EVF represents an initial state in the simulations of the long-term dynamical evolution. A reasonable assumption on the size of EVF could be made based on the results from the numerical simulations of asteroid disruptions. Still, more accurate data for a specific family needs to be extracted from what we know about this family. However, the EVFs are often determined from the long-term evolution simulations, similar to the family ages. Therefore their estimation is coupled with both the evolution process and age determination. It implies that, in principle, family age and EVF could be simultaneously determined by simulating the evolution of the family until it matches the available observational constraints. This is, however, tricky, and much better results could be obtained if age is somehow independently constrained.

Therefore, while modelling a post-impact evolution of an asteroid family allows, in essence, reconstruction of the family's initial ejection velocity field, its efficiency depends from case to case. This approach works the best for young asteroid families, though in some cases, reliable results could also be obtained for older families. 

A pioneering attempt in defining the size–ejection velocity relationship for asteroid families' members was made by \citet[][]{1999Icar..141...79C}. The authors found an inverse dependence of ejection velocities on the size, characterised by an exponent between $-2/3$ and $-1$. While successful in explaining the available data on asteroid families at that time, its important limitation was that it did not account for the family evolution. The result could be at least in part compromised by the fact that the Yarkovsky effect is causing a similar size-dependent evolution. Therefore, the identified size–ejection velocity relationships might be consequences of the post-impact evolution due to this effect rather than an outcome of the initial collision event. Even more sensitive to this analysis are the obtained ejection velocities, which may be overestimated by \citet[][]{1999Icar..141...79C}. However, we note that this analysis was done using objects larger than about 10~km in size, a population known at that time. These are large enough objects that, in some cases, their Yarkovsky-induced post-impact evolution should be relatively limited. Therefore, the result could be still valid to some degree.

Among the subsequent work, let us mention a series of papers by \citet[][]{2006Icar..182...92V, vetal2006,2006Icar..183..349V}. These authors reconstructed the post-impact evolution of several asteroid families and found that the initial EVFs were significantly smaller compared to the present-day situation.
In particular, a typical ejection velocity of 5~km fragments in the Erigone, Massalia, Merxia, Astrid, and Agnia families was found a few tens of meters per second. This value means the initial semi-major axis dispersal in analysed families was $25-50\%$ of the observed value. The post-impact evolution naturally explained such a situation, and newly derived initial velocities match well the results of hydrocode simulations and laboratory experiments ~\citep{michel15_asteroidIV}.

For results derived from young asteroid families, we refer interested readers to Sections~\ref{yf} and \ref{vyf}. However, let us recall here the results obtained by \citet[][]{karin2006} for the Karin asteroid family as those changed the paradigm of how large the initial velocity fields of asteroid families. The Karin was the first young asteroid family discovered. Being just 5.8~Myr old, the evolutionary mechanisms did not have enough time to modify Karin family fragments' velocity field significantly. Therefore, the members of the Karin family provide direct insight into its original EVF. 

\citet[][]{karin2006} performed a set of hydrocode simulations of the Karin family formation, and found good fits to the size-frequency distribution of the observed fragments and the ejection speeds inferred from their orbits. The results suggested that the ejection speeds of smaller fragments produced by the collision were faster than those of the larger fragments, with the mean values for $>3$-km-diameter fragments being about 10 m~s$^{-1}$. 

Despite this success, it was clear that precise determination of the initial EVFs is a complex issue and that, especially for older families, novel methodologies are needed to deal with these challenges. The problem was recently reassessed by \citet[][]{2016MNRAS.457.1332C}. The authors investigated the possibility to use only the vertical component $v_{\rm W}$ of the EVF, which is related to the orbital inclination. The motivation behind this approach is that the proper orbital inclination is generally less affected by the dynamical evolution processes than the orbital semi-major axis and eccentricity. \citet[][]{2016MNRAS.457.1332C} found that the distribution of the $v_{\rm W}$ should be more peaked than a Gaussian distribution (i.e. be leptokurtic) even if the initial distribution were Gaussian. They surveyed known asteroid families to distinguish such cases and identified eight families characterized by the leptokurtic $v_{\rm W}$ distribution. All these families are located in dynamically quiet regions of the main belt, and their initial orbital configuration should almost not be modified by subsequent orbital evolution. Therefore, these families are good candidates for the reconstruction of the initial EVF.

One of these eight identified families is the Koronis family, and in the follow-up work, \citet[][]{2016Icar..271...57C} investigated the initial velocities in this group. The main results are that the spread in the original ejection speeds is consistent with an inverse size-velocity interdependence, and the minimum ejection velocity is of the order of 50 m~s$^{-1}$. These results are fully consistent with results of \citet[][]{2006Icar..182...92V, vetal2006,2006Icar..183..349V}, and regarding the velocity-size relationship also with \citet[][]{1999Icar..141...79C}, further strengthening our knowledge about the EVFs in asteroid families.

A similar approach is afterwards applied to some other asteroid families. In addition to the parameters of the Astrid family described in Section~\ref{s:dynevo}, using the fact that the $v_W$ component of terminal ejection velocities is quite leptokurtic, \citet{2016MNRAS.461.1605C} has also estimated terminal ejection velocity parameter to be the range V$_{\rm ej}$ = 5$^{+17}_{-5}$ m~s$^{-1}$. The author excluded values of V$_{\rm ej}$ larger than 25 m~s$^{-1}$ from constraints from the current inclination distribution. Using similar arguments and methodology, \citet{2017MNRAS.465.4099C} found the terminal ejection velocity parameter of the Hoffmeister family. Based on the constraints from the observed inclination distribution, the authors concluded that the family V$_{EJ}$ should be lower than $25$~m~s$^{-1}$, with V$_{\rm ej}$ = $20\pm5$~m~s$^{-1}$ being the most likely value. Note that the obtained ejection velocity parameters match well the escape velocity from the parent bodies of these families. Additionally, independent verification of the size of the initial EVFs was obtained by \citet[][]{2018MNRAS.473.3949B}, based on the V-shape methodology.

The list of works on the initial velocity fields of asteroid families given here is by no means complete. It just intends to demonstrate the main challenges and results obtained along these lines, focusing primarily on the more recent papers. Nevertheless, these demonstrate that the past efforts allowed us to understand the typical sizes of the initial EVFs and a general size-velocity dependence. Especially, the inverse dependency of ejection velocities on fragment sizes has been verified, as well as the sizes of the initial EVFs, which approximately correspond to the escape velocities from the corresponding parent bodies.

We still do not understand well asymmetries in the EVFs, though available evidence suggests they are not uncommon outcomes of the disruption events. Just as an illustration, \citet[][]{2006Icar..182...92V} found that the velocity component normal to the Eos orbit plane is by a factor 4 larger than the along-track velocity component. A similar situation was also noted in the Veritas family \citep[see e.g.][]{nov2010}. While the theoretical motivation for asymmetric velocity fields is well justified \citep[see][]{1994Icar..107..255B}, their quantification for individual families is challenging. Additional works are also needed to understand better how the initial EVFs in asteroid families depend on material properties and the internal strength of parent bodies. Future works are expected to move the step further, synthesize the results into the global picture, and connect them with other relevant aspects such as velocity distribution among colliding asteroids or interior structures of asteroids.

\subsection{The Yarkovsky/YORP-based age determination}
\label{ss:yarko_ages}

An underlying fundamental issue for proper understanding of many problems that could benefit from the knowledge extracted from families is that apriori we do not know the ages of those groups. Dating the asteroid families thus comes as a critical topic, and once families are identified, we want to know the time of the initial collision that formed them. In principle, ideally, age is what we should also know to model the long-term evolution of an asteroid family. However, as discussed above, age determination is often coupled with modelling efforts. Still, different long-term evolution models present different levels of coupling (interdependence) with the age of the family. 

Several dating methods are developed and available, each with its own advantages and disadvantages. When dating asteroid families, we need first to distinguish between young (and very young) families formed less than 10~Myr ago, and those older than about 50~Myr. It is because orbits of members of young families could still be accurately tracked backwards until the time of the fragmentation event. On the other hand, older families do not offer that possibility, and considerably different dating methodologies need to be applied. As the young families are extensively discussed in Section~\ref{yf}, including their appropriate dating methods, this section focuses on the age determination of older families. We review here the Yarkovsky/YORP chronology-based dating methods, which are the most promising approach for older families. One of the method's main advantages is that it can be applied to almost all families older than 50 Myr, while providing relatively reliable results.

\emph{How does it work?} Asteroid families evolve over time, and this is the key to computing the time of the initial collision: the smaller pieces generated by the impact tend to change their initial configuration, moving far away from the parent body. If we can understand how far and how fast they moved over time, we can then find the age of the family.

The Yarkovsky effect is a non-gravitational perturbation produced by
the anisotropic thermal emission of thermal photons from the asteroid
surface. It depends on the object's physical properties
(e.g. thermal inertia, Bond albedo, density) and in particular, it is
proportional to the inverse of the diameter. As a result, its effect
is larger on smaller bodies \citep{vok2000, farnocchia13}, producing a
substantial drift of the semi-major axis over million of years. Due to the Yarkovsky effect, the smallest members of the family tend to move farther away from the parent body, creating the so-called V-shape.

\begin{figure}[h]%
\centering
\includegraphics[width=0.7\textwidth]{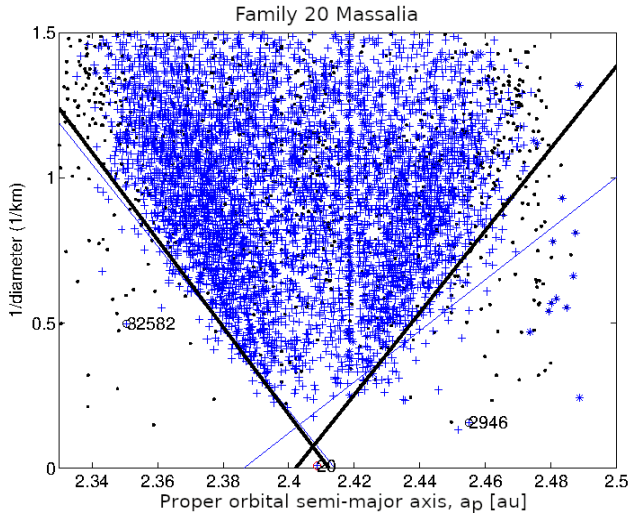}
\caption{"V-shape" of the family of (20) Massalia. The figure
  represents the family in the $(a_{\rm p},1/D)$ plane, where $a_{\rm p}$ is the proper semi-major axis in astronomical units, and $D$ is the diameter in km. Blue crosses identify the family members, while black
  dots are the background objects. The two solid black lines are
  the result of the fit to the sides of the V as
  in~\citep{2015Icar..257..275S}. The blue lines represent the different steps of the fit while trying to converge. The red circle around the number (20) gives the position of the parent body, while the black circles
  around blue crosses identify the position of outliers that the fit has rejected. The blue stars represent 3/1 resonant objects.}\label{fig:20vshape}
\end{figure}

This behaviour becomes clear when we plot the V-shape of
the family in the $(a_{\rm p},1/D)$ or $(a_{\rm p},H)$ plane, where $a_{\rm p}$ is the
proper semi-major axis, $H$ is the absolute magnitude, and $D$ is the
diameter of the family members (see Fig.~\ref{fig:20vshape}).

At the beginning of this century, \citet{2001Sci...294.1693B} set the stage for the Yarkovsky based chronology of asteroid families by recognizing the role of the effect in spreading families in terms of the orbital semi-major axis. 

\citet{netal2003} were the first to apply the Yarkovsky chronology-based method to estimate the age of an asteroid family.\footnote{The idea was for the first time presented a bit earlier by \citet{vok2002acm}.} In this work \citet{netal2003} used a simple approach by fitting an envelope to the distribution of the family in the $(a_{\rm p},H)$ space.
In this way, they were able to estimate the age of the Themis family to be $2.5\pm1.0$~Gyr. Later on, \citet{2005Icar..173..132N} used the same approach to estimate the ages of many other asteroid families.

A step forward in Yarkovsky chronology-based age estimation was taken in papers by \citet{2006Icar..182...92V,vetal2006}. In these works, instead of simply fitting an envelope in the $(a_{\rm p},H)$ plane, the authors used a Monte Carlo-like numerical approach. They fit several parameters simultaneously, including the initial velocity field of the family and the distribution of family members in terms of the semi-major axis. Observation of the coupled influence of the Yarkovsky and YORP effects was also a novel aspect of this work. This allowed the authors to obtain more robust age estimations for several families.

A generally similar, but yet somewhat improved, methodology to the one used by \citet{netal2003} was recently
proposed by \citet{2015aste.book..297N}. The authors opted to estimate the family ages based on the following equation:
\begin{equation}
T_{\rm age} \simeq 1~{\rm Gyr} \times \left( \frac{C}{10^{-4}~{\rm au}} \right) \left( \frac{a}{2.5~{\rm au}} \right)^2 \left( \frac{\rho}{2.5~ {\rm g~cm}^{-3}} \right) \left( \frac{0.2}{p_v} \right)^{\frac{1}{2}},
\label{eq:age_nes2015}
\end{equation}
where $\rho$ is the asteroid bulk density and $p_v$ is the visual geometric
albedo. The constant $C$ is related to the age of the family and depends on the semi-major axis drift.
It is defined as $C=10^{-H/5} \lvert a_{\rm p}-a_{{\rm p}, c} \rvert$, where $a_{{\rm p}, c}$ is a center of the family, and $H$ is an
absolute magnitude.

Another way to approach the problem has been proposed 
by~\citet{2015Icar..257..275S}. It is also based on the finding an envelope to the distribution of the family, but in the $(a_{\rm p},1/D)$ space, and this time the envelope is computed through a least-squares fit. An additional difference concerning described techniques is that the age is here estimated from the slopes of the V-shape, not from the position of the V-shape borders. We will explain soon why this matters,
but let us first explain the method in more detail.

The first step consists of finding how much the orbits of the smaller
members moved over time, i.e. how much the proper semi-major axis
changed. This is obtained through a linear regression fit applied
independently to the two sides of the V-shape:
\begin{itemize}
  \item The two sides of the V-shapes are sampled using bins of
    different sizes. The goal is to obtain bins with a similar
    density. Consequently, bins will be larger at the bottom and
    smaller at the top of the V-shape.
  \item For each bin, a representative is chosen. For the left (inner)
    side, it corresponds to the element with the lowest proper
    semi-major axis value. For the right (outer) side, it is the
    element in the bin with the largest $a_{\rm p}$. The two sets containing
    all the representatives can be seen as a sample of the inner and
    outer sides of the V.
  \item A linear fit is performed on each side, considering a
    weighting scheme and the outlier rejection as
    in~\citet{carpino03}. See an illustration shown in Fig.~\ref{fig:20vshape}.
\end{itemize}

The inverse of the slope obtained as a result of the fit represents the
space spanned by the family members due to the Yarkovsky effect that
we call $\Delta a$. This is true for both sides of the V-shape.

The second step of the method consists of finding the drift rate at
which the fragments moved over time. Since the Yarkovsky effect has
never been directly measured from observations for individual objects in the main belt, \citet{2015Icar..257..275S} rescaled the
value detected for the asteroid $(101955)$~Bennu \citep[e.g.,][]{chesley14,farno2021}, to
obtain reasonable measures for each family, according to their known
physical and spectral properties (note that the corresponding values for Bennu are accurately known). The formula used for the rescaling is:
\[
\frac{da}{dt} \simeq \frac{da}{dt}_{\rm Bennu}
\frac{\sqrt{a_{\rm Bennu}}(1-e^2_{\rm Bennu})}{\sqrt{a}\,(1-e^2)}
\frac{D_{\rm Bennu}}{D}\frac{\rho_{\rm Bennu}}{\rho}
\frac{\cos\gamma}{\cos\gamma_{\rm Bennu}}\frac{1-A}{1-A_{\rm Bennu}}
\]
where $\rho$ is the density chosen for each family, and $A$ the Bond albedo. It is also assumed that $\cos\gamma=\pm 1$ for obliquity, extremizing the effect, depending upon the inner or outer side of the V-shape. The age of the family is thus given by: $T_{\rm age} = \Delta a / (da/dt)$.

Generally speaking, however, the Yarkovsky effect, and therefore the time evolution of the V-shapes, may not necessarily scale as $1/D$. For instance, \citet{2007Icar..190..236D} suggested that thermal inertia depends on a diameter of an asteroid as $\Gamma = D^{\alpha-1}$. Based on this, \citet{2018A&A...611A..82B} proposed that Yarkovsky drift scales as $(da/dt) \propto D^{-\alpha}$, and consequently a size-dependent modification of V-shapes. However, \citeauthor{2018A&A...611A..82B} did not find evidence for this hypothesis among 25 investigated main-belt asteroid families.

In addition to the described cases, an interesting aspect of the Yarkovsky chronology is met for families whose members are locked inside the powerful mean-motion resonances. The resonances lock the semi-major axis in these cases, and the Yarkovsky-induced drift translates into eccentricity. This mechanism produces a V-shape in the $(e_{\rm p},1/D)$ plane, in an analogy to more typical V-shapes seen in the $(a_{\rm p},1/D)$ plane. By determining appropriate scaling factors, the V-shapes in the $(e_{\rm p},1/D)$ plane could also be used to determine the family ages \citep[][]{2008MNRAS.390..715B,milani17}.

Finally, let us recall here that the age of an asteroid family could be approximately estimated also based solely on the YORP effect.
The idea relies on the \emph{YORP-eye} concept introduced by \citet{2016Icar..274..314P}. The YORP-eye refers to a depopulated region in the V-shape plot of the family that could be diagnostic of the age of the family \citep{2019MNRAS.484.1815P}. More in particular, the position of this area within the V-shape depends on the age of the family. Assuming a non-dimensional parameter \emph{YORP-age} $Y_{\rm age}$ is simply given by the equation: $Y_{\rm age} = \tau_{\rm f} A / a^2$, where $\tau_{\rm f}$ is the age of the family (in Myr), $A$ is Bond albedo, and $a$ is the semi-major axis (in au).
Then, calibrating its value on families with known ages, the ages of other families could be estimated. We refer readers to the two previous papers for further details on the algorithm. The YORP-eye method could be useful to get an independent estimate of families ages. These estimates, however, are usually affected by large uncertainties and, often, not unique.

\emph{What are the good and bad sides of different versions of the Yarkovsky-based chronology? And what are their general limitations?}

The first thing that should be understood is that the V-shape is generally the result of two processes: the initial ejection velocity field and secular evolution of the semi-major axis due to the Yarkovsky effect. At first, larger fragments tend to be ejected at lower velocities and smaller fragments at higher speeds. Thus, the initial configuration
after the collision is expected to see the smaller fragments located further
away from the parent body ~\citep{michel15_asteroidIV}. After that, being more sensitive to the Yarkovsky effect than the larger ones, smaller fragments also move even farther away from the parent body. Separating the contribution of these two effects is a complex issue. Therefore, this, to some degree, affects the accuracy of all the variants of the method. The less sensitive and probably the most accurate should, in principle, be the method developed by \citet{2006Icar..182...92V,vetal2006}. Its main shortcoming is that it cannot be so easily applied to many families, but needs to be used on a case-by-case basis.

Both variants based on an envelope fitting, either in the  $(a_{\rm p},H)$ or $(a_{\rm p},1/D)$ plane, are affected by the contribution of the initial velocity field, which is not taken into account. An important aspect to realize here is that method used by \citet[][see also Section~\ref{ss:ini_vel_field}]{2015Icar..257..275S} depends on the initial velocity field differently from the classical Yarkovsky chronology-based age determination. The point is that \citeauthor{2015Icar..257..275S} determined ages from the slopes of the V-shapes and not from the position of the V-shape borders. The slopes do not generally depend on the initial size of the family, which is the case in classical V-shape envelope fitting methods. This, however, does not mean that the \citeauthor{2015Icar..257..275S} version of the method is independent of the initial velocity field. With the ejection velocities being correlated with the size of the fragments, the initial velocity field already produces a V-shape like distribution. For this reason, the borders of the V-shape are not vertical lines in the  $(a_{\rm p},1/D)$ plane, and this initial V-shape slope affects the final age estimate. 

Regardless of the method used, the contribution of the initial velocity fields to the current distribution of family members decreases with the ages of asteroid families. Therefore it should be less important for families older than ~1 Gyr. Still, the probability of family formation suggests that families formed from larger parent bodies are older \citep{b2005}. At the same time, ejection velocities are, in principle, also correlated with the size of the parent bodies. The larger bodies eject on average fragments at larger velocities. Therefore, the older families, which are typically also older, are expected to be formed in collision events with higher ejection velocities. Therefore, even for older families, the initial velocity field could play a role in age determination.

The other aspect may also affect the accuracy of the age estimation. The Yarkovsky calibration is one of the most limiting aspects. The effect depends on many physical properties such as density, surface thermal properties, spin state, etc., unknown for most asteroids. Another important mechanism is the YORP effect, which modifies the spin states of asteroids, determining the magnitude of the Yarkovsky effect. Finally, the dynamical effects such as close encounters with massive asteroids could randomize the semi-major axis to some degree, obscuring the V-shape distribution.

Having all these in mind, we can expect that current family age determinations are at best accurate on a factor of about 2.
This should be considered in any analysis based on the estimated family ages.

\subsubsection{Recent Family Age Estimates}\label{Subsec:results}

\citet[][]{2015Icar..257..275S, milani17} and \citet{2015aste.book..297N} provided age estimates, or necessary information to directly compute those ages, for more than 50 families. The methods used, along with their limitations, are discussed in Section~\ref{ss:yarko_ages}. The main strength of these estimates is that they are obtained using uniform methods, which is particularly important in reconstructing a collisional history of the main belt.

\begin{figure}[h]%
\centering
\includegraphics[width=0.48\textwidth,angle=-90]{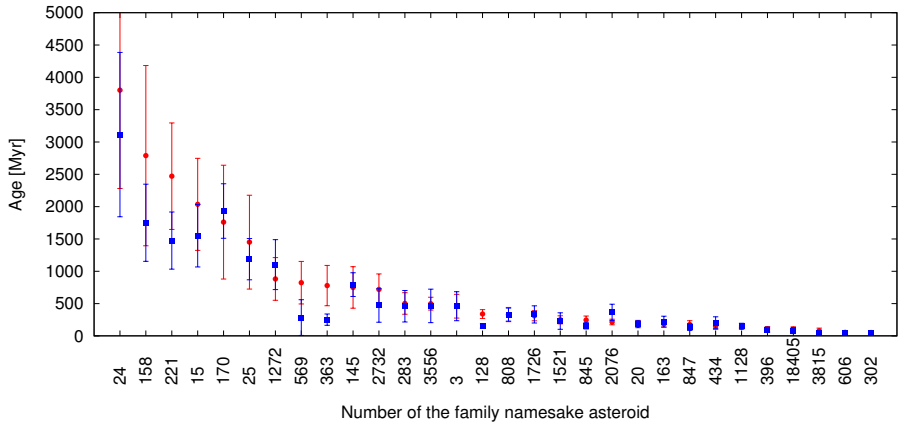}
\caption{Ages and their corresponding errorbars for 30 families estimated both by \citet{2015aste.book..297N} (red circles) and \citet{2015Icar..257..275S,milani17} (blue squares). The $x$-axis shows the number of the family namesake asteroid (e.g. 24 for Themis), while the $y$-axis represents the time span of millions of years. Families are sorted from the oldest (top left) to the youngest (bottom right).} \label{fig:chronology}
\end{figure}

The obtained results are shown in Fig.~\ref{fig:chronology}. The ages referring to \citet{2015aste.book..297N} are computed using Eq.~(\ref{eq:age_nes2015}) and information given in this work, and the same density as in \citet{2015Icar..257..275S}, to allow more direct comparison. There are a few important points to notice here. The results are generally in good agreement and compatible from the statistical point of view. In two cases where the error bars do not overlap (for families 569, 363) the disagreement is caused by different names used for the same family, or the same name for different families. For instance, (569)~Misa contains a younger subfamily, namely (15124)~2000 EZ39. While \citet{2015Icar..257..275S} under the age of (569)~Misa provides the result for the older family,
\citet{2015aste.book..297N} refers to the name (569)~Misa to the younger subfamily. Since the significant difference in the results. Similarly, the difference in age estimates for (363)~Padua should result from different memberships. However, the exact reason is not apparent, and this case might be worth further investigation.

In any case, most of the results from the two independent estimates are in good agreement. This suggests that the ages are relatively reliable. However, it should be noted that uncertainties are significant. Let us illustrate this in the (24)~Themis family. The method developed by \citet{2015Icar..257..275S} uses an independent fitting of both sides of the V-shape.
In the case of the Themis family, the authors obtained the age of $2450\pm840$~Myr from the inside slope and $3780\pm960$~Myr from the outside slope. Though formally statistically compatible, these two nominal results are not so close each to other. At the same time, data from \citet{2015aste.book..297N} suggests an age of $3800\pm1500$~Myr. Combining these three estimates, we got that the Themis family is $3340\pm1970$~Myr old. It is a much older solution than the usually quoted age of $2.5\pm1.0$~Gyr found by \citet{netal2003}. At the same time, uncertainty is huge. Hopefully, a better solution will be found shortly, as Themis is one of scientifically the most interesting asteroid families, with a space mission also being proposed \citep{2022P&SS..Themis}.

Having mentioned two independent age estimates from the in- and out-side slopes of the  V-shape by \citet{2015Icar..257..275S}, it is worth mentioning that in some cases, this yields different ages for the same dynamical family. Such a situation is encountered when the two sides of the V-shape have significantly different slopes, likely because the dynamical family results from multiple collisions. The best example is the asteroid (4) Vesta. Two different ages have been found for the family~\citep{2015Icar..257..275S}. They could be interpreted as two different families, corresponding to the two largest basins observed by the Dawn mission, namely Rhea Silvia and Veneneia. The estimated formal ages for the two craters are $1$ Gyr and $2$ Gyr respectively~\citep{marchi12}, in agreement with the two ages found for the family of (4) Vesta by~\citet{2015Icar..257..275S}.

We discussed here only age estimates from \citet{2015Icar..257..275S, milani17} and \citet{2015aste.book..297N}, as these works provided the results for many families using the uniform method. There are, however, many other age estimates, often given for a single or a few families. In many cases, such estimates should be of somewhat better accuracy than those obtained semi-automatically for a large number of families. It is beyond our scope to mention all these works. Let us, however, discuss one striking example that suggests even more caution is needed when interpreting family ages. \citet{2020A&A...643A..38Y} recently analysed (31)~Euphrosyne family and found an age of $280^{+180}_{-80}$~Myr. 
It is significantly younger than the estimate of $1230\pm410$~Myr coming from \citet{2015Icar..257..275S}. Previously, the age of the family was also estimated to $\sim500$~Myr, by modelling population of family members interacting with the $\nu_{6}$ secular resonance \citep{2014ApJ...792...46C}. Though in principle, one may expect those age estimations based on the evolution caused by gravitational interactions used by \citet{2020A&A...643A..38Y} and \citet{2014ApJ...792...46C} could somewhat underestimate the age, the differences are still immense and raise a warning.

From a more broad perspective, Fig.~\ref{fig:chronology} also points out the limitation of our current
knowledge of asteroid families:
\begin{itemize}
  \item There is an evident absence of primordial families in the plot. This bias appears because of the clustering method used in the family classification~\citep{2015aste.book..297N,deienno21}. The hierarchical clustering methods usually cannot identify very old families, because they are too dispersed in the space of orbital elements.
  \item The uncertainties are too large to obtain a reliable
    chronology, and ages overlap. These immense uncertainties come from a lack of knowledge of asteroids' physical and spectral properties and from the rescaling used to obtain reasonable values of the Yarkovsky effect in the main belt.
\end{itemize}

\subsubsection{Future age determination perspectives}

We expect significant progress in the precision of family age estimates in the future. On one side, discovering new asteroids and improvements in our knowledge of the physical properties of asteroids will be very useful to obtain more accurate ages. On the other hand, existing methods should be further improved and extensively tested. The testing should be practical for quantifying their uncertainties and identifying the primary sources of the errors. Regarding the physical properties, along with data expected from future sky surveys, an example of good practice is a step taken by \citet{sergeyev21} in mining old archival data to find new information. The authors searched for moving objects in the SkyMapper Southern Survey and extracted their photometry and astrometry from the images. The result extends the known multi-filter photometry and taxonomic classification for more than $200\,000$ objects. 

Recent work by \citet{2021MNRAS.506.4302D} is an example of steps that we need to follow to understand better the uncertainties and limitations of the family age estimation models. These authors investigated the role of physical processes on age estimations. Indeed, tests performed by \citet{2021MNRAS.506.4302D} are simplistic, focusing only on some of the relevant effects. The approach neglects the full effect of dynamical mechanisms or the YORP effect. At the same time, though not explicitly stated in the paper, a number of test particles used to represent synthetic families seem to be significantly larger than what we have in a realistic situation. Combined with neglected effects of resonances, it makes V-shape borders fitting unrealistically accurate. Nevertheless, this is a helpful first step, which needs to be improved further.

Regarding the calibration of the Yarkovsky effect, the ESA Gaia mission will represent a real game-changer in the estimation of the ages of asteroid families. Originally designed to
create the most accurate $3D$ map of the
galaxy~\citep{prusti16_gaia,brown18_gaia}, launched in December 2013,
the Gaia spacecraft also produces highly accurate observations of
solar system objects. The second release, in April 2018, showed for
the first time the improvements that Gaia is going to make to asteroid
orbit determination~\citep{spoto18_gaia}. The post-fit residuals of
Gaia asteroid observations reach the mas level for bright
asteroids, two orders of magnitude better than what we reach from
Earth using ground-based observatories. For the first time, we can
think about having the same accuracy for main-belt asteroids that we
obtain for near-Earth asteroids when we observe them with the
radar. This is crucial for detecting the Yarkovsky effect in
the main belt. This non-gravitational perturbation can be measured
from the astrometry using a non-linear least-squares fit, where the
effect is one of the parameters to be
solved-for~\citep{farnocchia13}. The more accurate the orbit, the
higher the probability of detecting the Yarkovsky effect on small
objects. Gaia finally opened this opportunity for the asteroids in the
main belt. The goal is to obtain at least one measurement for every
family, or at least every spectral type. That will help solve
the problem of large uncertainties in age determination.

\subsection{Families delivering fragments on Earth}
\label{ss:showers}

The fact that families play an important role in producing near-Earth asteroids and meteorite flux on the Earth is not new \citep[e.g.][]{1992CeMDA..54..207Z,1993Sci...260..186B,1995Icar..118..132M}. As families evolve over time, many fragments ecsape from them and are often not anymore recognized as members of these families \citep{2018NatAs...2..528N}. Some of the escaped objects are transported towards the near-Earth space. This topic has fascinated scientists already for about half a century. It is also a natural consequence of asteroid family members being a significant fraction of the whole population of main-belt objects. Over the years, we have learned a lot about the related processes, but nevertheless, there are still many open questions, which are among the central problems in asteroid science. We primarily need to understand: i) what is the overall contribution of each family? ii) how the flux from a family evolves with time? and iii) which specific NEAs (meteorites) originate in a given family?

But before answering the above questions, we need to know which families could contribute some members to the NEAs population. In principle, the answer is simple, all of them. However, there are two situations to distinguish regarding the time evolution of the flux from an individual family. Only families close to the main transport routes from the main belt to the near-Earth region could inject fragments directly onto those routes. The dynamical lifetimes of the objects located inside the strong orbital resonances, identified as the primary transport routes of NEAs, are very short, in many cases only a few Myr \citep[e.g.][]{granvik-etal_2018}. Therefore, these directly injected fragments could be transported quickly, causing many objects to reach the near-Earth region in a short time, soon after the formation of the family. Such disruption events in the main belt could cause variation in the flux towards the near-Earth region. \citet{1998Icar..134..176Z} suggested that the formation of the most populous families located close to the transport routes should have produced transient episodes of intense craterization of the terrestrial planets. The duration of these \emph{asteroid showers} depends on the resonance involved and on the number of injected fragments in each case. On the other hand, families that do not inject the fragment directly onto the transport routes send a more stable number of objects into the near-Earth region. Therefore, although all families could deliver fragments to the near-Earth region, only those close to the powerful resonances are also expected to cause \emph{asteroid showers} on Earth.

The motivation of \citet[][]{1998Icar..134..176Z}  was to explain how short-lived asteroid populations inside the powerful resonances are replenished with fresh fragments to keep the population of near-Earth objects in a steady state. It was, however, immediately apparent that the ejection velocities of the fragments in family-forming phenomena are too low to inject enough objects into resonances. The problem of bringing new objects into the resonances was solved soon afterwards by recognizing the importance of the Yarkovsky effect \citep[e.g.][]{1999Sci...283.1507F}. Nevertheless, the hypothesis of Zappal{\` a} et al. about the transient \emph{asteroid showers} turned out to be correct.

NEAs population models are all statistical representations of the ‘whereabouts’ of NEAs of different sizes. They are built by studying the influx rate of NEAs from different sources, estimating via dynamical simulations the mean residence time of incoming objects in different parts of near-Earth orbital space, and weighing the results against observations \citep[][]{bottke-etal_2002, granvik-etal_2018}. An implicit assumption behind these statistical representations is that the NEA population is close to a steady state, which may be true if/when temporal variations of influx are minor. However, the asteroid showers might not fit into this view, and imply that some family forming events in the main belt can break assumptions that the NEA population is in a steady state. But do we have evidence for such showers?

One of the ways to constrain the variation in the meteorite flux on Earth, is by studying the meteorites that we have in our possessions, as well as dust deposits on our planet.
Most meteorites that fall today are H and L type ordinary chondrites. However, the main-belt asteroids best positioned to deliver meteorites are LL chondrites \citep[][]{2013Icar..222..273D}. By elemental and oxygen-isotopic analyses, \citet[][]{2017NatAs...1E..35H} studied this contradiction and found that the current meteorite flux is dominated by fragments from recent asteroid breakup events and therefore is not representative over longer timescales. Additionally, Heck et al. found that the meteorite flux varied over geological time as asteroid disruptions create new fragments that slowly fade away from collisional and dynamical evolution. The current flux favours disruption events that are larger, younger, and/or highly efficient at delivering material to Earth.
Similarly, at the other end of the size spectrum, a recent micro­meteorites study found changes in the flux composition, possibly due to the direct delivery of a meteoroid swarm from the main asteroid belt to Earth \citep[][]{2019Geo....47..673D}. Additional evidence for impact flux variations comes from the work of \citet{2019Sci...363..253M} who studied the terrestrial impact crater record and found that the impact rate increased by a factor of 2.6 about 290 million years ago. All this evidence suggests that asteroid families play a major role in meteorite delivery to the Earth and that their formation may cause variation in the Earth’s impact flux.

Different aspects, including dynamical evolution and transport routes, size of the parent body, initial velocity field, and cosmic-ray exposure (CRE) ages, need to be considered to establish direct links between the asteroid showers and families. Here we will discuss some of the most engaging examples, focusing on more recent results.

\begin{figure*}
 \begin{center}
  \includegraphics[width=0.9\textwidth]{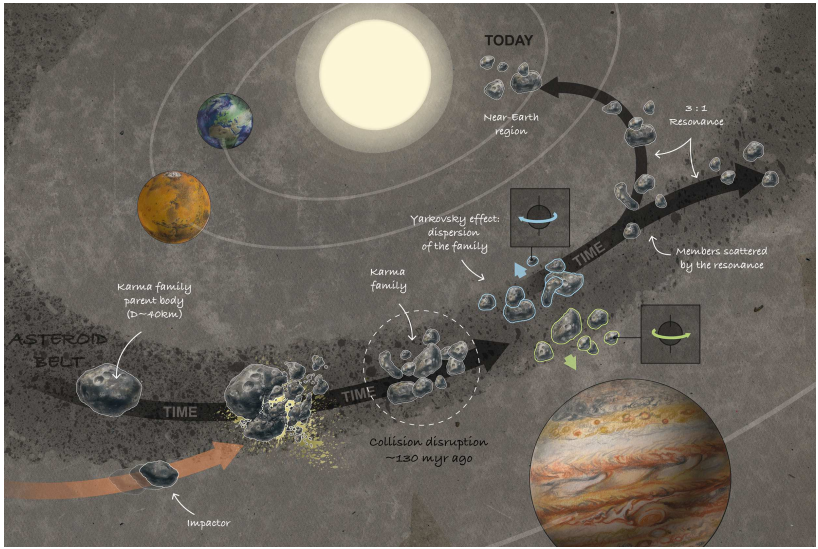}
 \end{center}
 \caption{An illustration of the main steps in a possible scenario in the dynamical evolution of an asteroid family. From a family formation to the delivery of members into near-Earth space. The example shows a scenario for the Karma asteroid family after the work by \citet[][]{2021MNRAS.501..356P}.}
\label{fig:karma}
\end{figure*}

One of the first dynamically based links between meteorites and their parent asteroids was proposed by \citet{2009Icar..200..698N}. The authors identified the Gefion asteroid family as a source of the shocked L-type chondrite meteorites, representing two-thirds of all L chondrite falls. To verify this connection, Nesvorny et al. followed two different lines of evidence. On one side, they searched for possible locations in the main belt capable of transporting meteorites to the Earth on timescales comparable with the most extreme CRE ages of the fossil meteorites \citep[$<$200 kyr;][]{2004Natur.430..323H}. On the other side, the age of marine limestone in southern Sweden, where the fossils of L-type chondrite meteorites and iridium enrichment were found, as well as the ages of five large terrestrial craters, are all found to be about 470 Myr. Because of these facts, and given the short transport timescale, the authors looked for a group of objects formed around 470 Myr ago. Nesvorn{\' y} et al. found that only the 5:2 mean motion resonance with Jupiter at 2.823 AU can transport the meteorites over the given short timescale and that the age of a nearby Gefion family plausibly coincides with the shock event of L chondrites dated 470~Myr. Therefore, the authors concluded that the Gefion asteroid family is likely the source of the fossil L chondrite meteorites.

The perfect example of how an asteroid disruption in the main belt may affect the near-Earth environment is the case of the Flora family. Being close to the $\nu_6$ secular resonance, the Flora family has been proposed as a significant source of a large number of NEOs and LL-type ordinary chondrites meteorites \citep[e.g.][and references therein]{2015aste.book..243B}. What, however, was not well constrained were the timing and magnitude of Flora family members striking the Earth. \citet{2017AJ....153..172V} recently re-accessed the Flora family and performed a very detailed study aiming to constrain better the time variations in the influx of near-Earth objects from the family.
The authors used state-of-the-art tools to simulate the long-term collisional and dynamical evolution of the family and investigated different assumptions about the initial dispersion of the fragments.
Though there are some uncertainties in the exact number of objects and the timing of their arrival in the near-Earth environment, \citet{2017AJ....153..172V} undoubtedly showed that the flux gradually decreased to the present-day situation after an initial spike of objects. Therefore, in an interval between about 100-200 Myr after its formation, the Flora family sent an increased number of fragments towards the Earth, likely causing an asteroid shower-like event. Such a scenario was possible only because many fragments produced in the Flora family forming event were injected directly into the $\nu_6$ resonance.

In Section~\ref{s:dynevo} we already discussed the work of \citet[][]{2015Icar..247..191B} in light of its dynamical simulation advances. However, in this paper, \citeauthor{2015Icar..247..191B} actually presented a systematic analysis of the transport from the five primitive inner main-belt asteroid families (Erigone, Sulamitis, Clarissa, Eulalia, and New Polana) to the near-Earth region. The authors primarily aimed to identify potential source families of asteroid Bennu, a target of NASA's OSIRIS-REx sample return mission. By tracking the evolution of millions of test particles started near the most likely main-belt source regions for Bennu, \citet[][]{2015Icar..247..191B} found that Bennu mostly likely arrived into the near-Earth region via the $\nu_6$ resonance. The intermediate source Mars-crossing region (IMC) is a possible but less likely source region, while the 3:1 resonance is very unlikely to transport an asteroid onto a Bennu-like orbit. The applied methodology
allowed \citeauthor{2015Icar..247..191B} to estimate that the relative probabilities of Bennu coming from the $\nu_6$  resonance and the IMC region is $82$\% and $18$\%. Using the Monte Carlo simulations of the semi-major axis drift, including also for the first time stochastic YORP model, it was possible to measure the flux of Bennu-like objects from each of five considered families into each relevant transport route. Combining the simulations results and the estimates of the probability of reaching a Bennu-like orbit via specific transport routes, the authors found that the New Polana and Eulalia families had about $70$\% and $30$\% probability of producing Bennu, respectively. Interestingly, the same methodology was applied to asteroid Ryugu, a target of JAXA's Hayabusa~2 sample return mission, and found similar source region probabilities as for a Bennu, opening a possibility that both objects originate from the same parent body.

In more general terms, it is important to underline here that the results obtained by \citet[][]{2015Icar..247..191B} are size-dependent, meaning that objects of different sizes placed on the same orbit as Bennu and Ryugu could have different families as the more probable source location. For instance, as noted by the authors, for bodies that are considerably smaller than Bennu, the flux from families like Erigone should start to play an increasingly important role.

A link has also been established between aubrite meteorites and the Hungaria family \citep[see e.g.][and references therein]{2020GeCoA.277..377G}. However, as shown by \citet[][]{2014Icar..239..154C}, the meteoroids from the Hungaria region predominantly reach Earth due to Yarkovsky induced drifting across the orbit of Mars, with no assistance from orbital resonances. This results in their longer ($\sim 50$ Myr) delivery times, in agreement with the CRE ages of aubrites. Therefore, though the Hungaria family definitely supplies meteorites to the Earth, it probably did not cause any transient shower of these objects.

An interesting example is also the Baptistina family. Based on the numerical simulations of the collisional and dynamical evolution of this group, \citet[][]{2007Natur.449...48B} proposed the Baptistina family as a source of Cretaceous/Tertiary (K/T) impactor. Despite an extensive simulation effort, at that time, the authors were missing a piece of critical information about the composition and density of the family members. It prevents them from adequately modelling the transport from the family to the Earth. Once the critical data about the properties of the Baptistina family became available, it was apparent that potential Earth's impactors, coming from this family, neither match the expected composition of the K/T impactor nor the timing of impacts coincide with this event \citep[][]{2009M&PS...44.1917R,2012ApJ...759...14M}. Nevertheless, although the Baptistina is likely not a source of K-T impactor, the work of \citet[][]{2007Natur.449...48B} still informs us that the family contributes some objects to the NEAs population. In particular, it could be an additional source of LL chondrite-like NEAs \citep[][]{2017AJ....153..172V}.

There are also other examples of families found to deliver some asteroids in the near-Earth region.
For instance, \citet{2015ApJ...809..179M} performed numerical simulations of the dynamical evolution of the Euphrosyne asteroid family. The family is one of the largest low albedo
groups known, and thanks to the large orbital inclinations of its members, it is crossed by the $\nu_6$ secular resonance. This provides a unique opportunity for an outer main-belt group to deliver fragments directly to the $\nu_6$ resonance and via the resonance to the near-Earth region.
Similarly, \citet{2021MNRAS.501..356P} found that the Karma family, a group of primitive asteroids located at the outer edge of the 3:1 MMR with Jupiter, which is about 140 Myr old, has been supplying some asteroids to the near-Earth region. Currently, there should be about ten family members larger than 1 km in diameter, and many more smaller objects, orbiting in near-Earth space. finally, \citet[][]{2017MNRAS.471.4820A} found that the Maria family could contribute some objects to the NEOs population, though the contribution of bodies larger than $~3$~km in size should be practically zero at present. 

The list of families sending objects towards the near-Earth region is certainly not complete. We discussed some of the most interesting examples and mentioned a few new findings. However, as stated above, any family could, in principle, contribute to the NEOs population. Therefore, many new links are expected to be established in the future.

\section{Special family classes}
\label{sec:special_families}

\subsection{Young families} \label{yf}

Denomination \emph{young} in the title of this section refers to a specific type of chronology, namely the situation when purely dynamical methods may infer the family age. The history of young families started with a
seminal paper by \citet{mf1994}, who studied a compact, outer main-belt
Veritas family. Analysing the past orbital history of its seven members, they argued that this family must have an age of about $50$~Myr or less,
because two of these members --including (490) Veritas-- moved away from the tight family zone on this timescale. It was the first example when chaotic dynamics were used to determine the age of a family. While fascinating by itself, the accuracy of the chaotic chronology has its apparent limitations. It would have been more promising if a chronology based on regular dynamics was used. A milestone in this direction was set by \citet{karin2002}, who reported a discovery of a compact family about the Koronis member (832) Karin, therefore the Karin family.
More importantly, noting the fairly regular nature of orbital evolution
in the Koronis zone, they used convergent behaviour of secular angles (aka backward integration method - BIM)
for $13$ orbits of Karin members to show that this family is $5.8\pm 0.2$~Myr old. Little later, \citet{netal2003} re-analysed the structure of the Veritas family in detail, finding that this case also offers a possibility of regular-orbit-based determination of its origin, aside from the method mentioned above based on chaotic orbits. This is because a certain portion of the phase space occupied by Veritas members holds
sufficiently regular orbits. \citet{netal2003} thus observed that past
nodal convergence of Veritas members having regular orbits implies
an age of $8.3\pm 0.5$~Myr. The Veritas and Karin families thus
became exemplary archetypes of young asteroid families. While similar
in this sense, they also represent quite different worlds, and we comment
on them in some detail below. The past two decades have seen the astronomy of young families flourishing, and we thus overview some of the most interesting results.\footnote{This section presents only some of the most fascinating and best-studied young families. There are also other known families younger than about 10~Myr that deserve further study.
For instance, \citet{2012MNRAS.425..338N} present a preliminary analysis of the Lorre cluster and found it was formed in a cratering event onto about 30~km large parent body about 1.9~Myr ago. More recently, \citet{2018MNRAS.479.4815C} used the backward integration method to study many potentially young families and found four of them to be less than 10~Myr old.
\citet{2019MNRAS.482.2612T} analysed the population of Eos family members and found its young subfamily, namely about 2.9~Myr old Zelima cluster. This cluster was later on studied by \citet{2020P&SS..18204810C} who used its resonant nature to set independent constraints on the age and initial ejection velocity field. Note also those young families connected with main-belt comets are presented in Section~\ref{ss:water}.}

\smallskip

\noindent{\it Karin family.-- }The discovery of the Karin family by \citet{karin2002} became an important landmark of planetary science in general. This is because
aside from recognising a new class of families that are interesting in their own right, it helped to solve several standing problems in planetary astronomy. Before we get to this context, we first review the main steps in understanding the Karin family itself.

Interestingly, each step in a better definition of the Karin family brought along new scientific results. The initial paper of \citet{karin2002} had $39$ members, of which $13$ were suitable for an experiment on the past convergence of the secular angles. The propagation model contained gravitational perturbations from the planets being the primary driver of orbital precession in space. The next step was taken by \citet{karin2004}, who made use of quickly increasing
new asteroid discoveries. They identified already $90$ Karin members, of which $34$ had suitable orbits to allow improvement of the family age to $5.75\pm 0.05$~Myr. The smaller uncertainty achieved was not the only result of a larger asteroid sample, but it also stemmed from the improvement of the dynamical model. Aside from the gravitational perturbations, it now contained thermal accelerations (the Yarkovsky effect) for each Karin member. Determination of the associated secular drift rate of the semi-major axis ${\dot a}_{\rm Y}$ 
was an important side product of the experiment because it allowed broadly constraining the surface properties of these small main-belt asteroids. The possibility to promptly identify family interlopers, otherwise a difficult task (see Section~\ref{ss:interlopers}), is an additional important aspect of the age-determination via past orbital convergence. This is because the interloper orbits do not follow the convergence pattern. As an example, \citet{karin2004} proved the originally proposed second-largest
Karin member --(4507) Petercollins-- does not belong to the family. It has massive importance in the interpretation and model reconstruction of the family size distribution. \citet{karin2016}, revisiting the Karin family identification, revealed yet another dynamical effect. By that time, they had already $576$ Karin members, out of which $480$ were suitable for repeating the past convergence method of \citet{karin2004}. The much larger sample of family members for which orbits converged had an immediate result of improving the accuracy of Karin's age: $5.75\pm 0.01$~Myr. More importantly, though, \citet{karin2016} also estimated and analysed the semi-major drift-rates ${\dot a}_{\rm Y}$ due to the Yarkovsky effect for the $480$ Karin members with sizes in
between $1$ and nearly $6$ kilometres. The values of ${\dot a}_{\rm Y}$ for $55$ asteroids with sizes in between $2.5$ and $3.5$~km peaked at zero value and extended symmetrically to some maximum negative and positive values compatible with theoretical expectations. They were consistent with a sample of objects having an isotropic distribution of spin axes.
However, the ${\dot a}_{\rm Y}$ values for $280$ asteroids with sizes
between $0.9$ and $1.7$~km told a different story. Their distribution was also roughly symmetrical, but had maxima at some negative and positive values with only a very few objects having ${\dot a}_{\rm Y}\simeq 0$.
\citet{karin2016} showed that this distribution is compatible with a sample of objects whose initially isotropic pole distribution was tilted by the YORP effect toward ecliptic poles (thus extreme values of the obliquity).
Therefore the analysis of the Karin family origin brings evidence of not
only the Yarkovsky effect, but also the YORP effect. There is another
important context for this result. Note that the chronology of moderately old families proposed by \citet{vetal2006} makes use of the same synergy of the Yarkovsky-YORP effects to reproduce the polarisation of small members in the family toward extreme values of the semi-major axis. Here the same dynamical processes are caught in action at an earlier phase, when the effect on the semi-major axis is still small. However, the bimodal obliquity distribution is already apparent.

Given the pristine state of the Karin family, observed only a few Myr after its formation, it became a primary target suitable for comparison with results from numerical simulations of collisional fragmentation of large asteroids. Only some of the best-characterised very young asteroid families (Sec.~\ref{vyf}) may rival Karin's exclusive status by now. So far, the most detailed work was presented by \citet{karin2006}. These authors found that the Karin family was formed by an impact of $\simeq 5.8$~km projectile onto $\simeq 33$~km parent body with $\simeq 6-7$ km~s$^{-1}$ speed and $\simeq 45^\circ$ impact angle.
The largest created remnant, (832)~Karin, has about $17$~km size, and the size distribution of Karin fragments then follows from the second largest fragment of $\simeq 5.5$~km with a steep cumulative power law of $\simeq -5.3$ exponent extending to about $2$~km sizes. Modelling thus indicates that Karin resulted from a rather energetic collision with $\simeq (0.1-0.15)$ mass ratio between the largest remaining fragment and the parent body. As to the ejection velocities, \citet{karin2006} found that $\simeq 3$~km fragments (typical to the steep leg in the size distribution) have mean barycentric ejection speeds of $\simeq 12$
m~s$^{-1}$ with the fastest (and smallest) launched at $\simeq 30$ m~s$^{-1}$. These values are comparable, or only slightly larger, than the escape speed from the parent asteroid, namely $\simeq 20$ m~s$^{-1}$. While fairly complete and satisfactory, \citet{karin2006} pointed to several unresolved issues that need a more complete study in the future (especially if combined with a new update of Karin's population). For instance, predicting the right order of magnitude of the ejection velocities of Karin members from the largest remnant, their exact 3D velocity field was not successfully reproduced. Related to this issue may also be the fact that the model so far has failed to reproduce spin parameters of the largest remnants: (i) (832) Karin is a slow prograde rotator with the period
of $18.35$~hr \citep[e.g.,][]{sm2012}, (ii) while the next six to seven large
fragments are retrograde rotators (some of which also have rotation periods determined and found to be shorter than that of Karin). Matching such a configuration is still a challenge for impact modelling.
The spin state of (832) Karin itself may be a sort of curiosity,
if not a mystery, as it bears similarity with Slivan states of larger and Gyr-old members in the Koronis family \citep[e.g.,][]{vetal2003}. However, before going too far along with this speculation, one first needs to break the longitudinal ambiguity of Karin's pole through future observations.

While expectedly very useful for the physics of collisional disruptions,
discovering the Karin and Veritas families, identifying their properties and determination of their young age, became crucial in solving the following decade-long problem in planetary science. Analysing all-sky scans in thermal waveband taken by IRAS spacecraft, \citet{low1984} discovered warm, large-scale structures (bands) extending symmetrically up to $10^\circ$ away from the ecliptic plane. The inferred temperatures of $150-200$~K pointed to their origin in the asteroid belt, most likely dust released in asteroid collisions. A more complete description of these interesting structures was later presented by \citet{s1988}, and confirmed by observations of the COBE spacecraft \citep[e.g.,][]{r1997,k1998}. Soon after their discovery, sources for the dust in these bands were passionately debated. \citet{der1984}, observing that the latitudes of the most prominent $\alpha$, $\beta$ and $\gamma$ bands roughly
coincided with the mean proper inclinations of the prominent Themis, Koronis and Eos families, proposed that these giant and old families as the sound sources. A competing model, in which the dust in the bands is produced in Myr-old collisions of moderately large asteroids ($15-20$~km or so), was formulated by \citet{sg1986}. While the former hypothesis was initially preferred, support for the latter view stemmed from about a degree mismatch between the Eos proper inclination and the latitude of the $\gamma$ band, and by realising that the age of Themis, Koronis or Eos families is much larger than the timescale to reach collisional equilibrium with the whole main belt for their dust component. So the traces of their dust should be presumably smeared into a smooth signal from the whole belt. \citet{karin2002} and \citet{netal2003} immediately recognised the value of the young age of the Karin and Veritas families in the context of the discussion on the IRAS bands origin. These sources perfectly fit what \citet{sg1986} had in mind. In fact, the proper inclination of the Veritas family matched very well with the $\beta$ band latitude, and removed thus the pending problem with Eos. A detailed numerical model of the scenario, in which Karin and Veritas were the sources of the $\beta$ and $\gamma$ bands, was developed by 
\citet{bands2006}. Of the many results in this paper, we mention here an
important quantitative justification. \citet{bands2006} showed that the observed absolute flux from the $\beta$ band could be explained by the total volume of Karin-released $1\,\mu$m to $1$~cm particles equivalent to a sphere of $11$~km diameter (and similarly for Veritas and the signal from the $\gamma$ band). This comfortably fits in the above-mentioned disruption model of the Karin family parent body. The $<10$~Myr old Beagle family was little later found to be the source of the $\alpha$ band by \cite{2008ApJ...679L.143N} \citep[it is possible that the Beagle family is slightly older, or holds substructures, e.g.,][but the connection to the $\alpha$ band would not be changed]{2019P&SS..166...90C}. Strong support for the young-families origin of the IRAS dust bands also comes from an independent analysis of \citet{fetal2006}, who found traces of the Veritas-released dust in $^3$He anomaly detected in the globe-wide geologic layers of the appropriate age.

The second broad-scale topic in planetary science to which discovery of
Karin (and other young families) significantly contributed has to do with
space weathering. This phenomenon, known since the analysis of
the Apollo-era lunar samples, aims --in the asteroid context-- to resolve
a discrepancy between the observed asteroid reflectance spectra and
those of the presumed meteorite analogues (such as the S-type class
of asteroids and the ordinary chondrites). It has been suggested that the
asteroid reflectance spectra are being modified over time due to irradiation by cosmic rays, impingement by micrometeoroids and solar wind particles, their combination or other yet to be identified processes (see reviews by \citet{wea2002} or \citet{chap2004}). This was known by the end of the millennium. However, the characteristic timescale of the weathering processes, which may also have the power to discriminate between different variants of them, was not constrained by astronomical observations. The first attempt in this direction was provided by \citet{jed2004}. These authors used a compilation of ages for known S-type asteroid families and observed their correlation with the average spectral slope from broad-band photometry of their members.
To make the analysis robust enough, it was crucial to compare the data
for Gyr-old families (such as Maria, Eunomia or Koronis) to young
families (such as Karin and Nele/Ianinni, see \citet{netal2003} or 
\citet{cetal2018}; later \citet{ver2009} extended the sample to even
very young families, Sec.~\ref{vyf}). The extrapolation of the
colour-age correlation pointed rightly to the ordinary chondrite values even beyond the value observed for the young families. This implies
very fast onset of a weathered surface, while the observed slope of
the age-colour correlation described the slow tail of the processes
ongoing for Gyrs.

Could we anticipate what comes next? The asteroid database
keeps extending with discoveries of small asteroids, and this trend may
even accelerate in the future due to output from powerful sky surveys.
Hence, it is perfectly possible that the known Karin population could grow
to a few thousand in the next few years. Observing the trend described
above, the Karin age could be readily improved. More importantly, though,
the completion of the Karin population down to a few hundred-meter sizes would allow testing if the size distribution keeps increasing along
the power-law specified by large members \citep{karin2006}, or if a
slope change occurs. Because a $500$~m size asteroid
should have a collisional lifetime of about $200$~Myr \citep[e.g.,][]{b2005}, in little more than $5$~Myr of Karin's age, only about 3\% of the fragment population of this size should be affected. So if a pronounced slope change is observed above this size limit for Karin fragments, it would be more likely related to the fragmentation physics of the parent body. It would be exciting to see what the collisional codes predict at this level of resolution. However, there might be an
unexpected problem to deal with when so many new asteroids are
discovered. Recall that the primary identification method of the Karin
family is based on the hierarchical clustering tool in the
three-dimensional space of proper orbital elements. With the
main-belt population reaching a million asteroids or more, the
proper element space becomes very crowded. Already the Karin
family identification, despite its compactness, suffers from
overlap with the Koronis-2 family \citep[a recent cratering event on
(158) Koronis; e.g.,][]{mh2009}. Separating the two families may become
a delicate issue in the future. Perhaps the past orbital convergence
may be the key tool, but it is yet to be tested to what size limit
it would provide reliable results.
\smallskip

\noindent{\it Veritas family.-- }The Veritas family analysis by \citet{mf1994} brought a novel point of view on age determination, but it was not very constraining. A follow-up
work of \citet{kp2002} used many more orbits of Veritas members and
was more detailed in analysis, but in fact, resulted in an even looser age constraint of less than $100$~Myr. A masterful step in using
the chaotic chronology was thus presented by \citet{tsi2007}. These
authors first looked in detail at the orbital space occupied by the Veritas family, and using Lyapunov times associated with each orbit in this zone, \citeauthor{tsi2007} identified main zones of interest: (i) two chaos-revealing groups A and B, characterised by a short Lyapunov timescale between $10$ and $50$~kyr, and (ii) two regions R1 and R2 containing much more stable orbits. Importantly, the two largest remnants in the family are located in the opposite camps: (490) Veritas in size, the most chaotic group~A, (1086) Nata in the regular region R1
\citep[see already][]{netal2003}. Focusing mainly on the methods of the chaotic chronology, \citet{tsi2007} analysed in detail dynamics in zones A and B, which are related to the three-body mean motion resonances 5J-2S-2 and a higher-order 3J+3S-2. They observed a spread of the group-A and group-B Veritas members in proper eccentricity and inclination values. \citeauthor{tsi2007} found them larger than would have been expected by extrapolation from the adjacent regular zones R1 and R2. The extension was naturally interpreted as the chaotic diffusion
in the respective resonance over time. Thus, by conducting a numerical
experiment in which orbits are started in the expected initial region
inside each of the resonances, one can let the time go until the
spread of the synthetic orbital population matches the observed
population of the Veritas members. Using an elegant one-dimensional
diffusion model, in which the fundamental role is played by the
respective action invariant of the two-body heliocentric orbit,
\citet{tsi2007} proved that the group-A dynamics needs a timescale of
$8.7\pm 1.7$~Myr \citep[result that was later even improved using
the same method to $8.7\pm 1.2$~Myr by][]{nov2010}. This was in a 
marvellous agreement with the Veritas age determined from dynamics of 
regular orbits in the region R1 by \citet{netal2003}. Trouble, though,
occurred with the dynamics of the group-B members. Because the associated 3J+3S-2 resonance is much weaker, an order of magnitude longer timescale is needed to explain their spread in proper eccentricity. This problem lasts till now. The most likely solution consists of a spurious association of background objects to the family by the
clustering method in proper element space. In fact, this may partly
apply to the objects in the group A. Intriguingly, (490) Veritas
itself was proposed to possibly be such an interloper, as attempts to include it as the largest remnant in numerical reconstructions
of the Veritas family size distribution did not lead to satisfactory
results \citep[e.g.,][]{m2011}.

As noted above, the possibility to use regular dynamics for Veritas
family age determination was first discovered by \citet{netal2003}.
\citet{tsi2007} and \citet{nov2010} also used this possibility as
a comparison case to their chaotic chronology efforts. The novelty
was that they used now also orbits from the R2 region, and \citet{nov2010} the newly defined R3 region (a tail of the Veritas family containing orbits with semi-major axes larger than group A members in the 5J-2S-2 resonance). However, all these works used convergence of the orbital nodes only, deeming the much faster precessing perihelia unreliable. A full-frontal approach to using regular dynamics for Veritas age determination was finally taken by \citet{cetal2017}. Out of their 705 orbits in the R1, R2 and R3 regular zones, they selected 274 most stable ones, and successfully showed their simultaneous convergence in both the longitude of node and perihelion. Inspired by the Karin family analysis, they used the Yarkovsky semi-major axis drift rates ${\dot a}_{\rm Y}$ as a free parameter for each of the integrated orbits. This surplus of free parameters in their numerical experiment was the key element of their success. However, they also verified that the required ${\dot a}_{\rm Y}$ has the expected order of magnitude. The resulting Veritas family age was estimated to be 
$8.23^{+0.37}_{-0.31}$~Myr. Despite these advancements, the ``complicated world'' of the Veritas family keeps holding mysteries \citep[e.g.,][]{n2012}.

\smallskip

\subsection{Very young families} \label{vyf}
Very young families are compact clusters of asteroids sharing similar
heliocentric orbits with ages younger than $1-1.5$~Myr. Such ages imply that the values of secular orbital angles, the longitude of node $\Omega$ and perihelion $\varpi$, still have similar values (as well as semi-major axis, eccentricity and inclination). This is because the gradient of their proper frequencies $s$ and $g$ with a semi-major axis is typically several tens of arcseconds per year and astronomical unit. Given the initial ejection velocities of the order of ten meters per second in these structures, the fragments are initially dispersed in a zone of about $(1-2)\times 10^{-3}$~au in the semi-major axis near the largest remnant. So we roughly estimate that in a Myr, the secular angles disperse a couple of tens of degrees at maximum. Note that this value nearly always entirely dominates the initial secular angles dispersal by the
ejection velocity field, which is not larger than a fraction of a degree
\cite[see, e.g., Eq. (1) in][for a more quantitative discussion in the
specific case of the Datura family]{datura2017}. The fact that very young
families appear as clusters of asteroids in five-dimensional space of
osculating or mean orbital elements (excluding mean longitude) conveniently alleviates the problem of phase space filling with
an ever-increasing number of known asteroids. These structures appear as a cluster in the space of proper elements, but with a small number of known members (sometimes only three or four), their statistical significance may be poor. Adding two more dimensions represented by the secular angles helps isolate them from the background population of asteroids in most cases. Therefore the primary arena in which very young families are searched is the above-described five-dimensional space of osculating or mean orbital elements. In fact, the usefulness of proper orbital elements, in this case, is questionable (or limited at least) because the age of these families is often much shorter than the timescale over which the synthetic proper elements are constructed. Their values may thus average out some family structures of interest.

While many clustered structures, potentially very young families, have been reported in the literature, here we overview only those cases studied in sufficient detail. It may mean they contain many fragments, were the subject of targeted astronomical observations, or are of some particular interest.
\smallskip

\begin{figure*}
 \begin{center}
  \includegraphics[width=0.6\textwidth]{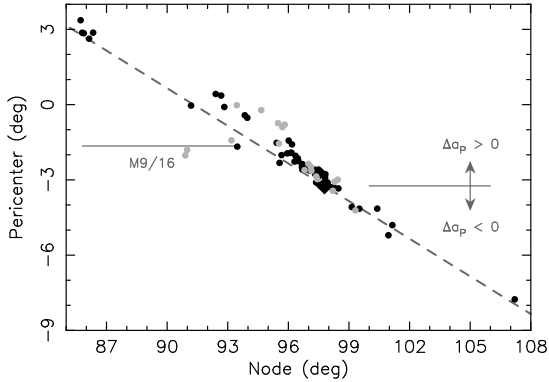}
 \end{center}
 \caption{Correlation of osculating values of secular angles,
  the longitude of node $\Omega$ at the abscissa and the longitude of
  perihelion $\varpi$ at the ordinate (at MJD 59400 epoch), for
  currently known population of $76$ members in the Datura family:
  black symbols are the $57$ multi-opposition asteroids (Datura itself
  shown by a diamond symbol), and $19$ single-opposition asteroids are
  shown using grey symbols. Members in the exterior M9/16 resonance
  have the nodal values between $\simeq 87^\circ$ and $\simeq 93.5^\circ$
  nodal longitude (as indicated by the interval at the left), members with
  proper semi-major axis smaller/larger than (1270)~Datura indicated by
  arrows on the right). The dashed line approximates the correlation
  using a linear function with a slope of $-0.5$. The largest dispersal
  around this trend is seen for objects affected by the M9/16 resonance
  (and some single-opposition objects whose orbits are still inaccurate).}
\label{f1}
\end{figure*}

\noindent{\it Datura family.-- }Historically, a compact cluster of six small asteroids about (1270) Datura (thence the Datura family) was the first example of a very young family discovered \citep{datura2006}. Its convenient location in the innermost part of the main belt, and large orbital eccentricity of $\simeq 0.21$ (thence low perihelia), allowed to quickly increase Datura's
known population to the present-day census of $76$ \citep[see gradual steps in][]{datura2009, rp2017,datura2017}. Fig.~\ref{f1} illustrates the novel aspect of the very young families in the case of Datura, namely the confinement of the secular angles to a limited interval of values quite smaller than the full range of $360^\circ$. The characteristic signature of the very young families is also the correlated nature of the secular angle distribution. The correlation is linear, as shown in Fig.~\ref{f1},
when a simple shear of secular angles due to the semi-major axis dependence of their precession frequencies, and their slightly different initial positions in the family, dominates the effect. As a result, fragments initially starting with a slightly smaller semi-major axis than (1270)~Datura are on Fig.~\ref{f1} driven to the bottom right corner and vice versa. Similarly to the young families with regular-enough orbits (such as the Karin family), the age-determination of the family is basically a reversal of
the same process, namely a numerical experiment of backward orbital propagation in which the currently observed configuration of secular angles (such as in Fig.~\ref{f1}) is shrunk to the expected initial state. Selecting a subset of dynamically stable orbits and the largest fragments in the family (the least affected by the thermal accelerations), \citet{datura2009} obtained $530\pm 20$~kyr for Datura
\citep[see also][]{datura2017}.

\begin{figure*}
 \begin{center}
  \begin{tabular}{cc}
  \includegraphics[width=0.95\textwidth]{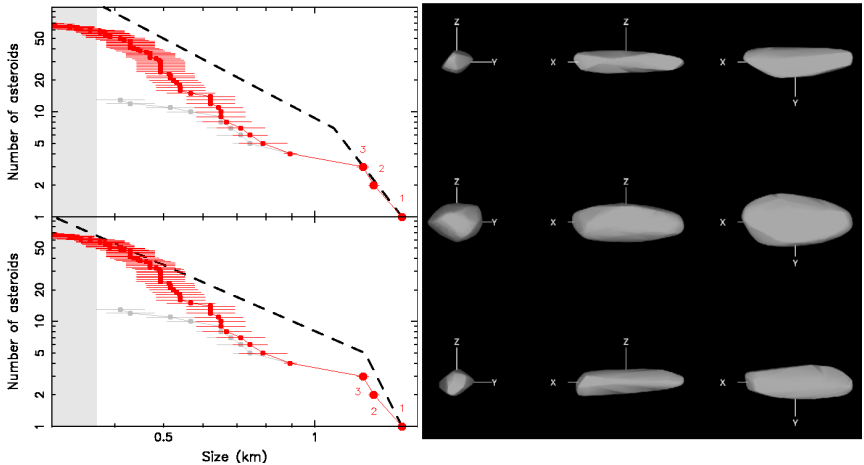}  
  \end{tabular}
 \end{center}
 \caption{Left panel: Size distribution of fragments in the Datura family (the
  largest remnant (1270)~Datura excluded): (i) the grey symbols are $13$ Datura
  asteroids detected by Catalina Sky Survey stations from 2005 to 2012, (ii) dashed
  lines show two possible solutions for a debiased population of Datura
  fragments \citep[assuming a simple broken power law, see][]{datura2017},
  and (iii) the red symbols are the presently known total population of
  $76$ fragments as of March 2022. Right panel: Shape models for the
  three largest Datura fragments based on optical photometry inversion:
  (i) (90265) 2003~CL5 on the top (labelled 1 on the left panel),
  (ii) (60151) 1999~UZ6 in the middle (labelled 2 on the left panel), and
  (iii) (89309) 2001~VN36 at the bottom (labelled 3 on the left panel).
  The left two views are perpendicular from the equatorial plane, the right
  view is pole on.}
\label{f2}
\end{figure*}

The convenient location of the Datura family, and the interest rose by
discovering its very young age in 2006, allowed and motivated several dedicated astronomical observations. In this sense, it has an exquisite status among the groups of very young families to date. 

Early photometric observations of (1270) Datura were obtained by \citet{w1997} and \citet{sze2005}, and allowed us to determine its short rotation period of $3.35$~hr. A dedicated campaign in the late 2000th provided enough additional photometric data to resolve the Datura rotation pole at $\simeq 76^\circ$ ecliptic latitude and a convex shape model \citep[see][]{datura2009}. Many more photometric observations of smaller fragments in the Datura family were
taken in the past decade. Those concerning the three largest of them
are interesting (Fig.~\ref{f2}): (90265) 2003~CL5, (60151) 1999~UZ6, and (89309) 2001~VN36 have all slow rotational periods of $23.41$~hr, $13.88$~hr and $73.15$~hr, respectively, their rotational poles have all ecliptic latitude larger than $55^\circ$, and their shapes are very elongated \citep[see][]{datura2017}. Because the lightcurve inversion methods allow for reconstructing only a convex hull of the shapes,  some, or all, of them may be contact binaries. On top of that, \citet{datura2017} used data about detections of $13$ Datura fragments by Catalina Sky Survey stations from 2005 to 2012, attempting to provide a debiased population down to about $200$~m in size. While the number of objects was still small, and thus the solution could not be strongly constrained, they found at least a range of possibilities (assuming a broken power-law size distribution). For instance, they predicted a population of $45\pm 12$ $D\geq 500$~m fragments, while currently, as of March 2022, there are about $30$ of them known. More importantly, the recent years' impressive growth of the known Datura population (red symbols at the left panel of Fig.~\ref{f2}) seems to exclude some of the shallowest predicted populations (such as the one shown using the hashed line at the bottom and left part of Fig.~\ref{f2}). The trend seen in the present data may also suggest that the two-slope, power-law model of \citet{datura2017} might be too simple and should perhaps be replaced with more complex parametrizations in future efforts. Overall, the information about the Datura family population, both shapes and rotation states of the largest members, and the size distribution downward the few hundred-meter sizes provides a unique challenge for outcomes from numerical simulations of asteroid collisions 
\citep[see, e.g.,][for some initial work]{hh2018}.

Spectroscopic observations provided additional information. For instance, \citet{mdn2008} found a progression from S to Sq and Q type spectra as we move from (1270)~Datura, about $9$~km size largest remnant in the family, down to kilometre-size small fragments. A more extensive spectroscopic study putting the data in the context of the space weathering processes was then published by \citet{ver2009}.
The fact that the space-weathering spectral slope indicator was not much different for the Datura family (and a few other very young families) and the young Karin family implies a rapid timescale for space weathering onset (possibly related to solar wind implantation in the surface matrix).

Datura family's very young age may qualify it well as a possible source of observable interplanetary dust (the same as much larger Karin, Veritas and Beagle families; Sec.\ref{yf}). \citet{vetal2008} attempted to model both orbital and collisional dynamics of dust released during the Datura family formation. Indeed, they found that Datura may potentially be a source of C and D dust trails observed by the IRAS spacecraft \citep[e.g.,][]{s1988}. Further confirmation using both modelling and independent observations will be needed in any case. In fact, \citet{datura2006} imagined more spectacular evidence of the very-young-families dust, namely its direct detection in the Antarctic and Greenland ice core drilling (after it accreted onto the Earth). Later estimates for Datura and other known families, however, indicated that the amount of possibly accreted dust is much smaller than would have
been needed.

Finally, despite the compactness of the Datura family, the orbital history of some of its members happens to interact with a weak, high-order, mean motion resonance M9/16 with Mars \citep[see already][]{datura2006,nv2006}. Various aspects and illustrations of the resonant dynamics in M9/16 has been studied by \citet{rp2017} and more recently by \citet{pr2021}. In general, such orbits present dynamical chaos effects that undermine their use for the family age determination based on regular-orbits convergence. The resonances crossing the very young families, such as the Datura family, are, however, too weak to allow chaotic chronology for independent verification of their age (such as in the Veritas family; Sec.~\ref{yf}).
\smallskip

\noindent{\it Schulhof family.-- }Another aspect that occasionally could turn a very young family analysis into a detective story, can be illustrated in the case of the Schulhof family. \citet{pv2009} discovered a small cluster of four asteroids about the largest body (81337) 2000~GP37 as a side product of their systematic search for asteroid pairs. Soon afterwards, \citet{vetal2011} realized that this is only a sub-cluster of a larger structure related to the largest remnant (2384) Schulhof, counting eight members altogether. 
(81337) 2000~GP37 with its group resides slightly offset from (2384) Schulhof in the orbital space. Such a situation perhaps requires special initial ejection velocity field properties, possibly related to the geometry of the initial, family-forming impact onto the parent object. Nevertheless, 
their common orbital convergence to (2384) Schulhof some $780\pm 100$~kyr ago justified a union of the two clusters of asteroids. However, \citet{vetal2011} opened Pandora's box by noting that two asteroids, tightly accompanying (2384) Schulhof in its newly discovered cluster, tend to orbitally converge less than $100$~kyr. They speculated that the first, larger-scale family-forming event had left (2384) Schulhof in an unstable state, facilitating the formation of a later secondary sub-cluster.

While later resurrected by \citet{fetal2020}, and perhaps best exemplified
in the case of the Emilkowalski family, the idea was retracted by a more
thorough work of \citet{vetal2016}. These authors found yet another four new members in the family (the current census as of November 2021 is $26$), but more importantly, they proved that the previously suggested young ages for some members were caused by their inaccurate orbits. The take-away message is that one has to carefully choose suitable members used for past orbital-convergence experiments, avoiding not only the chaos-affected resonant cases, but also those residing on single-opposition, or otherwise loosely constrained, orbits.
\smallskip

\noindent{\it Adelaide family.-- }\citet{ade2019} reported a discovery of a new, very compact cluster about the inner main-belt asteroid (525) Adelaide. Observing convergence of nodal longitudes, they suggested an age of $\simeq 500$~kyr. More recently, \citet{vetal2021} revisited the case of the Adelaide family for two reasons: (i) its hinted age appeared similar to that of the Datura family, and (ii) the two families are very close to each other in space of proper elements (consider the values $a_{\rm p}=2.2347$~au, $e_{\rm p}=0.1535$ and $\sin I_{\rm p}=0.0920$ for (1270) Datura and $a_{\rm p}=2.2452$~au, $e_{\rm p}=0.1487$ and $\sin I_{\rm p}=0.1170$
for (525) Adelaide). \citet{vetal2021} thus investigated a possibility of a
genetic relation between the two families, such that fragments created in the formation-event of one would trigger the formation of the other. First, considering a subset of Adelaide members residing on suitable orbits, they confirmed the Adelaide family age of $536\pm 12$~kyr. However, \citeauthor{vetal2021} found the case of a causal relationship between the formation events for the Datura and Adelaide families unjustified. In all likelihood, the two families were formed by an impact of a background-population projectile and, by chance, happen to be close to each other in time and orbital space.
\begin{figure*}
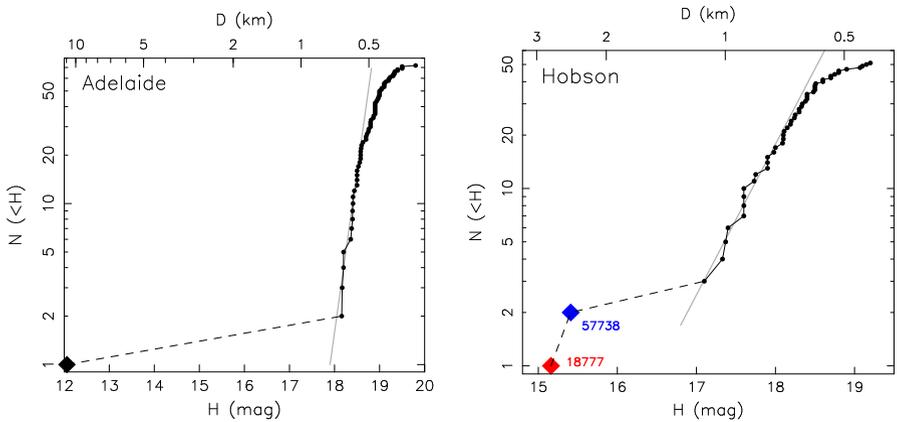

 \begin{center}
  \begin{tabular}{cc}
   \includegraphics[width=0.475\textwidth]{adelaide_sfd.ps} &
   \includegraphics[width=0.475\textwidth]{hobson_sfd.ps} \\   
  \end{tabular}
 \end{center}
 \caption{Left panel: Cumulative absolute magnitude distribution of the identified $72$ Adelaide family members. The largest remnant, (525) Adelaide, is shown by a diamond symbol, and small fragments by circles. The upper abscissa converts the absolute magnitudes $H$ to size $D$ using geometric albedo $p_v=0.22$, determined for (525) Adelaide by observations provided by WISE spacecraft. The grey line,
  approximating the small-fragment distribution, is a power-law $N(<H)\propto 10^{\gamma H}$ with $\gamma=2$. This is among the steepest progressions
  ever observed in collisionally-born families. Right panel: Cumulative absolute magnitude distribution of the identified $51$ Hobson family members. The largest objects, (18777) Hobson and (57738) 2001~UZ160 are highlighted by red
  and blue diamonds, small fragments by circles. The upper abscissa converts the absolute magnitudes $H$ to size $D$ using geometric albedo $p_v=0.2$, appropriate to the S-type classification of both largest asteroids. The grey line,approximating the small-fragment distribution, is a power-law $N(<H)\propto 10^{\gamma H}$ with $\gamma=0.85$.}
\label{f3}
\end{figure*}

Nevertheless, the analysis of \citet{vetal2021} pointed out yet another interesting aspect. The left panel in Fig.~\ref{f3} shows the cumulative magnitude distribution of the Adelaide family members: (525)~Adelaide appears to be about $10$~km size largest remnant in the family, accompanied by a large number of sub-kilometre fragments. While
it is still possible that some intermediate-size fragments are missing due to observational biases, the chances are not too high \cite[see, e.g., population completeness study of][]{hm2020}. Approximating the small-fragments tail using a power-law $N(<H)\propto 10^{\gamma H}$, we have $\gamma\simeq 2$. Assuming constant albedo value, this translates to a cumulative size distribution $N(>D)\propto D^\beta$, with
$\beta\simeq -10$. Either of the coefficients witnesses an extremely steeply-raising population \citep[compare, e.g., with relevant simulations of][]{setal2019}. Undoubtedly, the Adelaide family results from a huge cratering event on (525) Adelaide itself. However, the size-distribution steepness of the fragments is quite larger than observed
among cratering events of large main-belt bodies \citep[e.g.,][]{metal2013}. Hence again, a good example of motivation for impact-cratering modellers to test the codes, and relevant parameter space, in the regime of not too strong gravity. An interesting implication of the small-fragment population steepness is that the known population of 
the Adelaide family is among the most rapidly increasing over the past period. For instance, data published by \citet{vetal2021} had $52$ recognized members as of February~2021, while the present-day count is $72$ (March~2022; Fig.~\ref{f3}).
\smallskip

\noindent{\it Hobson family.-- }\citet{pv2009} reported a small cluster of four asteroids about the middle main-belt object (57738) 2001~UZ160, possibly a very young family with an age of less than $500$~kyr. They also noted a nearby asteroid (18777)
Hobson, but they were puzzled about its association with the cluster because of its similar size to (57738) 2001~UZ160. Later on \citet{rp2016,rp2017,rp2018} in a series of papers clarified this issue by proving that (18777) Hobson orbit converges very well to the members of the 2001~UZ160 cluster and must be considered a part of it. Because of its slightly larger size, the cluster was renamed the Hobson family. Additionally, by 2018 Rosaev and Pl\'avalov\'a updated the membership to nine members, noted a slight chaoticity of some orbits in this cluster (including that of 18777 Hobson, possibly due to the influence of a weak, three-body resonance 9J-8S-2), and used a simple past convergence method to infer an age of $365\pm 67$~kyr. This age was independently confirmed by \citet{petal2018}, who also updated the Hobson family population to $11$ objects.

The puzzling issue of a similar-size largest remnants in the Hobson family was reanalysed in a new light by \citet{hob2021}. First, these authors identified asteroids associated with this family from the most updated orbital catalogue. \citeauthor{hob2021} found the population increased dramatically (they had $45$ members as of July 2021, a number that even increased to the total of $51$ by the current date). They noted the extreme compactness of the Hobson family in the osculating values of their secular angles, with only $\simeq 1^\circ$ dispersion in longitude of node and $\simeq 2^\circ$ dispersion in longitude of perihelion (compare with data for the Datura family shown in Fig.~\ref{f1}). Therefore, the family membership is statistically robust, and all identified asteroids exhibit excellent orbital convergence among each other some $330$~kyr ago.
The cumulative absolute magnitude distribution $N(<H)$ of the resulting sample is shown in the right panel of Fig.~\ref{f3}. The conjoint existence of nearly equal-size largest remnants, followed by a disjoint population of smaller fragments, is not seen in the case of traditional asteroid families \citep[or can be easily excluded as an interloper situation, e.g.,][]{metal2013}. It has also not been predicted by computer simulations of asteroid collisions so far \citep[see][for relevant parent-body size]{setal2017,setal2019}.

By extending the efforts of \citet{setal2019}, \citet{hob2021} in fact, found a collisional regime that allows reproducing the observed size distribution of asteroids in the Hobson family. Nevertheless, the impact parameters must be fine-tuned, which decreases the statistical weight of such a solution. However, there is yet another interesting possibility. Observing results from \citet{petal2016}, who characterised the abundant population
of binaries among $\leq 15$~km size main-belt asteroids, \citet{hob2021} suggested a scenario in which the parent body of the Hobson family would be one of these binary systems. In this model, either the primary or the secondary undergoes collisional disruption, leaving the second component intact. The largest collisional remnant, and the former intact component in the binary, are thus the two large objects in the Hobson family. Because some $15\pm 4$\% of small main-belt asteroids are
binaries \citep{petal2016}, we statistically see more such configurations when a larger sample of very young asteroid families are discovered in the future.

\begin{figure*}
 \begin{center}
  \includegraphics[width=0.95\textwidth]{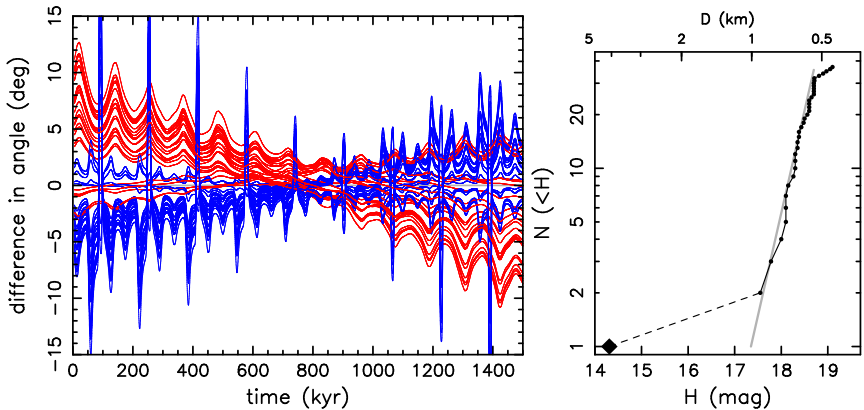}
 \end{center}
 \caption{Left panel: Illustration of convergence of the mean values of secular
 angles, the longitude of node $\Omega$ in red and longitude of perihelion $\varpi$ in blue, for a  selection of $25$ Rampo family members. The time at abscissa goes to the past, and the ordinate shows the difference of the respective angle with respect to the orbit of (10321) Rampo. Mutual convergence at $\simeq 800$~kyr is apparent. Right panel: Cumulative absolute magnitude distribution of the identified
  $37$ Rampo family members. The largest remnant, (10321) Rampo shown by a
  diamond symbol, small fragments by circles. The upper abscissa converts
  the absolute magnitudes $H$ to size $D$ using geometric albedo $p_V=0.2$,
  appropriate for the inferred S-type spectral classification. The grey line,
  approximating the small-fragment distribution, is a power-law $N(<H)\propto
  10^{\gamma H}$ with $\gamma=1.15$.}
\label{f4}
\end{figure*}

Excellent convergence of Hobson-member orbits in the past also provides an interesting constraint of the characteristic dispersal velocity of the observed fragments at origin, namely $\leq 10$~m/s. This is a very low value, comparable with the escape velocity from the proposed systems in both above-mentioned models, and well-reproduced in their simulations.
\smallskip

\noindent{\it Rampo family.-- }In order to demonstrate that quite more is at hand already at this moment, we give short information about the Rampo cluster. Its core, consisting of two small asteroids near the inner main-belt object (10321) Rampo, has been discovered by \citet{pv2009}. Being interested
in asteroid pairs rather than clusters, these authors did not pay much
attention to it other than noting a probable age between $0.5$ and 
$1.1$~Myr. More recently, \citet{petal2018} found another four Rampo members and improved the age using past mutual convergence of their secular angles to $780^{+130}_{-90}$~kyr.

Here we revisited the status of the Rampo cluster by searching in the up-to-date catalogue {\tt MPCORB.DAT} of asteroid orbits provided by the Minor Planet Center (we used the version as of March 2022). We found $37$ asteroids, including (10321) Rampo, associated with the Rampo family. Only one orbit is single-opposition, but some of the multi-opposition orbits also have only a few observations, and they are not very accurate (though enough for the association with the family). This represents a great population leap compared to status in 2018, indicating
that the past few years had a significant increase in known asteroid population propagated to the population status of several very young families. The left panel of Fig.~\ref{f4} illustrates the past convergence of mean values of secular angles for a selection of $25$ accurately determined orbits. We only used nominal orbits and a simple model in which only gravitational perturbations from planets, Ceres, Vesta and Pallas were included. All orbits tend to converge some $800$~kyr ago, in good agreement with results in \citet{petal2018}. However, more quantitative results would need to use more intense numerical effort by considering (i) clone variants of the propagated orbits (reflecting initial orbital uncertainty), and (ii) thermal accelerations in the propagation model. The right panel in Fig.~\ref{f4} shows the absolute magnitude distribution of Rampo-family members. It appears to be well compatible with simulations of the collisional breakup of a single parent body, in this case, \citep[e.g.,][]{setal2017,setal2019}.
\smallskip

\noindent{\it More on the dashboard.-- }The previous examples of very young asteroid families showed situations in which the population of known members swung from few to many in the past few years. This population increase typically brings along 
the discovery of some unexpected and interesting features. Since this trend is expected to continue, perhaps even accelerate, in the coming years due to observations of powerful sky surveys, a lot of exciting asteroid science may be expected in this field. We finish this section by commenting on three population-scale recent results or trends.

First, \citet{petal2018} brought up an interesting conjecture about the formation process of very young asteroid families. The traditional asteroid families are exclusively formed by collisional disruption of a parent object, once existing on a similar heliocentric orbit in the main belt. Realizing that the characteristic size of the largest remnants in the known very young asteroid families is $D<15$~km, and observing that it overlaps with
the size of the parent body of asteroid pairs \citep[e.g.,][]{vn2008,pv2009,pra2010, petal2019}, \citet{petal2018} investigated whether their formation processes may also be the same. Since the asteroid pairs have been convincingly shown to form by rotational fission of the parent body prevalently \citep[e.g.,][]{pra2010,petal2019},
the question is whether all, or at least some, very young families form the same way. Using the same approach, namely correlation between the rotational period of the largest remnant in the family versus the cumulative mass of smaller fragments, \citet{petal2018} argued that the very young families share many similarities with the pairs. This would support their common formation mechanism. While very interesting, this result needs further investigation. In some cases, small-fragment populations might have been underestimated in this study due to their observational biases. A clear example is the above-mentioned example of the Rampo family. In other situations, the rotational-fission formation may hold very well. Consider, for instance, the small cluster consisting of two small asteroids about (21509) Lucascavin discovered already by \citet{nv2006} which still has the same population today. Either no more fragments exist in this cluster, or they are so small that their contribution to the total fragment size will be negligible.

Second, \citet{fetal2020} revisited the issue of multiple formation events in the very young asteroid families (postulated, but then retracted, in the case of the Schulhof family discussed above). Indeed, they found four candidate examples in which possibly two events in recent history might have launched different sets of fragments associated with the family. The most interesting is the case of the Emilkowalski family. \citet{nv2006} discovered that the middle main-belt asteroid (14627) Emilkowalski is accompanied by two small fragments forming thus a young cluster. Convergence of secular angles hinted at the age of $220\pm 30$~kyr, although the orbit of one of the two fragments --(126761) 2002~DW10-- indicated a mismatch in the convergence. The very young age of $\simeq 200$~kyr is, however, a perfect match in a broader context set by analysis of the solar system dust bands discovered by
the IRAS spacecraft \citep[e.g.][]{s1988}. Indeed, a careful analysis of the
$17^\circ$ M/N bands by \citet{espy2009} and \citet{espy2015} indicated they are incomplete, caught in the process of full formation \citep[see, also,][]{vetal2008}. This constrains the possible age to less than $\simeq 300$~kyr, and the proper orbital inclination $17.2^\circ$ of (14627) Emilkowalski is just needed. However, a more complete analysis of \citet{fetal2020} revealed a more complex structure: (i) two of the confirmed Emilkowalski members --(224559) 2005~WU178 and (256124)
2006~UK337-- indeed converge to the largest remnant in the family some $250$~kyr ago, but (ii) other four fragments --including (126761) 2002~DW10-- tend to converge approximately $1500$~kyr ago (with large uncertainty, though). So were there two cluster-formation events associated with (14627) Emilkowalski, or are some incorrectly modelled effects (including possibly the orbital uncertainties) fooling the results? This interesting problem would need to be revisited when more family
members are found. A broader-context hypothesis has been recently published by \citet{car2020a}, who argued that asteroid clusters that originated in rotational fission of a parent body are more frequently found in young families rather than old ones. While interesting, this conjecture also awaits further, more detailed analysis. Note, for instance, that the most logical test case --the Karin family (Sec.~\ref{yf})-- does not seem to support these conclusions \citep[see][]{car2020c}.

Third, we mentioned above that the known asteroid population had seen a great increase over the past few years. As a result, some compact clusters with few members now have tens of them. The same reasoning may mean that some of the previously found asteroid pairs \citep[e.g.,][]{vn2008,m2010,petal2019}, configurations of just two asteroids sharing very similar heliocentric orbits, may soon move to the category of clusters \citep[in fact][expected that some of the discovered pairs might be just a tip of the iceberg, hiding an asteroid cluster below the observational limit]{vn2008}. Examples could be already seen, but a more systematic search awaits for the future. For instance, \citet{petal2019} noted that the asteroid pair (4765) Wasserburg and (350716) 2001~XO105 could be accompanied by a third body 2016~GL253. In fact, the present-day catalogue of asteroid orbits reveals that the Wasserburg cluster contains already six members, all showing very good past orbital convergence. In the same way, the young asteroid pair (5026) Martes and 2005~WW113 has a third companion in 2010~TB155. Altogether, they form a three-object Martes cluster, which is presently the youngest in its category \citep[having an age of $18\pm 1$~kyr;][]{petal2019}.

\subsection{Water Content in Asteroid Families}
\label{ss:water}

Our view about the water-ice content possibly still present in the main asteroid belt has changed dramatically over the last decade or so. The asteroids are mostly considered dry objects, in contrast to the comets, which are expected to contain a significant amount of water-ice. This view started to change when \citet[][]{2006Sci...312..561H} presented data showing the existence of a population of comets in the main asteroid belt. These objects, coined main-belt comets (MBCs), have both the orbital characteristics of asteroids and the physical characteristics of comets. MBCs are today a subpopulation of the so-called active asteroids, which are asteroids that at least occasionally display a comet-like appearance. Different possible mechanisms are initiating such manifestation of the active asteroids, such as dust ejection due to impacts or rotational fission. However, the activity of MBCs is driven by the sublimation of volatile ices, implying that these objects contain some amount of water-ice.

Spectroscopic evidence of hydration is also common among primitive asteroids, with the most direct evidence being the presence of a strong absorption feature around the 3~$\mu m$ band. This feature corresponds to different modes of OH groups in hydrated minerals and H$_2$O molecules on the surface of the small bodies \citep{2015aste.book...65R}. Based on the shape of the 3~$\mu m$ band, the asteroids exhibiting this feature could be divided into several categories corresponding to the prototype body. The first category (\emph{sharp} or \emph{Pallas-like}) exhibits a sharp 3~$\mu m$ feature, attributed to hydrated minerals. The second group (\emph{Ceres-like}) exhibits a 3~$\mu m$ feature with a band centre at around 3.05~$\mu m$, superimposed on a broader absorption feature from 2.8 to 3.7~$\mu m$. The third group (\emph{Europa-like}) exhibits a 3~$\mu m$ feature with a band centre at 3.15~$\mu m$. The fourth group (\emph{rounded} or \emph{Themis-like}) is characterized by a rounded 3~$\mu m$ band \citep{2012Icar..219..641T,2015aste.book...65R}. All these are, to some degree, hydrated objects, with the fourth category being explicitly linked to the presence of H$_2$O ice. The representative members of the \emph{rounded} 3~$\mu m$ feature are asteroids (24)~Themis and (65)~Cybele \citep{2012Icar..219..641T}, with a similar band shape also found in members of Themis, Hygiea, and Ursula asteroid families \citep{2020A&A...643A.102D}.

\emph{Where are the families in this story?} The asteroid families are generally homogeneous in composition, and therefore the chemical properties of members of an asteroid family are expected to be similar. It implies that if one member contains water-ice, this is likely the case for the other members. Interestingly, the first asteroid with surface ice detected, namely (24)~Themis, is a member of an asteroid family. Similarly, available spectra of the family members show that most of them have a rounded 3~$\mu m$ feature indicative of water-ice. Most of the known MBCs also belong to families.

Therefore, the families are deeply relevant not only for understanding the origin of MBCs, but also for the origin and evolution of the water-ice in the main asteroid belt. As soon as it was recognized that MBCs typically belong to the families, links between the two populations started to be investigated. A possible connection between the MBCs and young asteroid families was first proposed by \citet[][]{2008ApJ...679L.143N}. The motivation to link MBCs to young families was based on the expectation that ice could survive at heliocentric distances $<3.3$~au over the age of the solar system, only if deeply buried inside the larger asteroids. In this respect, the formation of young families provided a natural explanation of how the deeply buried ice has been brought close to the surface. In years following the work by \citet[][]{2008ApJ...679L.143N}, a couple of other connections between the young families and MBCs have been proposed \citep[e.g.,][]{2012MNRAS.424.1432N,2014Icar..231..300N}.

As the available data has grown over the years, \citet{2018AJ....155...96H} performed the first comprehensive analysis of the links between MBCs and asteroid families. This work found a strong indication that MBCs are primarily members of primitive asteroid families, regardless of how old these families are. Such results suggest, on one side, that parent bodies of primitive families were typically water-bearing asteroids. On the other side, these findings imply that ice could survive in main-belt asteroids over the age of the solar system under certain conditions, in agreement with estimates of the buried ice loss rate \citep{2008ApJ...682..697S,2020Icar..34813865S}. Therefore, the young age of families hosting MBCs may not be critical, as initially suggested. Nevertheless, primitive young families are still plausible sources of MBCs. The more recent associations of MBCs to families further support this view. \citet{2018RNAAS...2..129N} found that 427P/ATLAS (2017 S5) is a member of the Theobalda family, while P/2017~S8 (PANSTARRS) is the first MBC associated with the Pallas family, and P/2019~A7 (PANSTARRS) is the first MBC known to belong to the Luthera family \citep[][]{2019EPSC...13..892N}. Therefore, the evidence supporting the link between the MBCs and primitive asteroid families is growing, and it is now very robust.\footnote{For more general links between active asteroids and asteroid families, we refer readers to \citet[][]{2018AJ....155...96H}.}

While we still do not understand the links between the families and MBCs properly, three families stand out as the main reservoirs of MBCs. These are Themis, Lixiaohua and Theobalda families. These three families are of very different ages, but their members all share primitive composition characteristics. An important step forward in fully solving the mystery of MBCs will be a better characterisation of these three families, particularly the Theobalda family, which has been poorly studied so far. This is a key to understanding why these families are reservoirs of MBCs, while some other primitive families are not. In what follows, we review the main facts about the Themis, Lixiaohua and Theobalda families and discuss some other families expected to represent a birthplace for some MBCs, though still not recognised as such.

\subsubsection{Main properties of three families associated with the most MBCs}

\noindent{\it Themis family.-- } The family is one of the first five identified, the so-called Hirayama families \citep[][]{1918AJ.....31..185H}. It is a prominent group of asteroids located in the outer main belt, with low inclinations and low eccentricities. The Themis family is dominated by compositionally primitive, C-type asteroids, with many of them exhibiting spectra suggestive of aqueously altered mineralogy, similar to carbonaceous chondrite meteorites \citep[][]{1999A&AS..134..463F,2016Icar..269...62L}. The family mean geometric albedo based on the data from \citet[][]{2016PDSS..247.....M} is 0.068, while recently \citet[][]{2021AJ....162...40J} found that the average thermal inertia of family members is about 40~J~m$^{-2}$~s$^{-1/2}$~K$^{-1}$.

The Themis family parent body is believed to be differentiated icy-asteroid, about 400 km in diameter \citep[][]{2010GeoRL..3710202C,2016A&A...586A..15M}, which was catastrophically disrupted roughly about 2.5-3.5~Gyr \citep[][see also Section~\ref{Subsec:results}]{netal2003,2015Icar..257..275S}. Observations and modelling have implied that asteroid (24)~Themis is the core of an icy planetesimal, and therefore the family members are likely also icy objects. The evidence supporting this hypothesis includes a near-infrared absorption feature attributed to water ice frost was detected in spectra of both, asteroid (24)~Themis \citep[][]{2010Natur.464.1320C,2010Natur.464.1322R} and another prominent member of the family, (90)~Antiope \citep[][]{2015Icar..254..150H}.

The population of active asteroids includes at least three Themis family members: 133P/Elst-Pizarro, 176P/LINEAR, and 288P/(300163) 2006~VW$_{139}$. A fourth active asteroid, 238P/Read, is considered a possible former Themis family member whose orbit has dynamically evolved to the point at which it cannot be directly associated with the family \citep[][]{2009M&PS...44.1863H}. All these four objects are likely MBCs based on the results of their dust modellings and recurrent activities \citep[see][and references therein]{2018AJ....155...96H}. Additionally, the proper orbital elements of a newly discovered active asteroid (248370) 2005~QN$_{173}$ \citep[][]{2021arXiv210914822H}, which is currently inside the 11:5 MMR with Jupiter, are also compatible with the Themis family membership, though a cause of its activity is yet to be confirmed.

The Themis family also includes two known younger subfamilies, the Beagle and 288P families. Both subfamilies have been dated within the past 10~Myr \citep[][]{2008ApJ...679L.143N,2012MNRAS.424.1432N}, but more recently \citet[][]{2019P&SS..166...90C} found that the Beagle family could be somewhat older (about 35~Myr). In addition to the Themis family, two MBCs have explicitly been linked to young families, 133P/Elst-Pizzaro to the Beagle group and 288P to the small cluster named after it.

While there are many ice-rich objects beyond the orbit of Jupiter, the Themis family is likely the largest reservoir of icy objects in the asteroid belt. As such, it is an important keystone for understanding how icy asteroids form and evolve. For that reasons, \citet[][]{2021BAAS...53d.180L} recommend the asteroid (24) Themis and its family members as prime targets for further study and suggest both telescopic and especially spacecraft campaigns be used to explore these bodies in the next decade. 

Because the Themis family is next to the 2:1 MMR with Jupiter, it is believed that some family members may have been captured and scattered by the resonance since the family formation \citep[][]{1995Icar..118..132M}. Recent work by \citet[][]{2020AJ....159..179H} suggests that there could be Themis family objects evolving onto Jupiter Family Comets-like orbits in the present day. Similarly, a fraction of family members has potentially reached the near-Earth region, and recent work by \citet[][]{2020Icar..34813865S} suggests that some amount of water-ice could survive such transport, representing possibly water-bearing near-Earth objects.
\smallskip

\noindent{\it Lixiaohua family.-- } Like the Themis, the Lixiaohua is also a primitive asteroid family located in the outer main belt. The Lixiaohua is, however, somewhat smaller and significantly younger than the Themis family. The Lixiaohua family was formed by a super-catastrophic disruption of up to 200~km large parent body \citep[][]{2007Icar..186..498D}, that had happen 155$\pm$36 Myr ago \citep[][]{nov2010}.

The family mean geometric albedo is $p_v=0.044$ \citep[][]{masiero15_asteroidIV}, which is about 2\% smaller in absolute terms than in the case of the Themis family. The spectroscopic properties of the Lixiaohua family are recently studied by \citet[][]{2020Icar..33813473D}. They found that a majority of the objects belong to primitive classes, such as the C- and X-complex, and T- and D-types. In particular, the two largest members, asteroids (3330) Gantrisch and (3556) Lixiaohua, are classified as T-type objects, which is an intermediate class between X and D-types.

\citet[][]{2020Icar..33813473D} also found that Lixiaohua has distinct properties from the Themis family (see also Section~\ref{sss:future:mbc}), which can be interpreted as a difference in composition or differentiation level. An important dissimilarity between these two groups is related to the 0.7~$\mu m$ hydration band. While the presence of this feature in members of the Themis family is well established, it is absent among the Lixiaohua family members. 

Despite the differences with respect to the Themis family and lack of identification of aqueous alteration traces in spectra of its members, two MBCs (313P and 358P) are robustly associated with the Lixiaohua family. To better understand the family and its links to the MBCs, further analyses are needed. In particular, observations at the 3~$\mu m$ region would be helpful to clarify if hydrated minerals are indeed not present in family members, as suggested based on the 0.7~$\mu m$ band.
\smallskip

\noindent{\it Theobalda family.-- } This is the youngest and smallest of three families with two or more MBCs associated. The Theobalda family was formed by an impact cratering on the largest fragment, asteroid (778)~Theobalda, that is found to occur about $7\pm2$~Myr ago \citep[][]{2010MNRAS.407.1477N}. The author also found that the family originates from a parent body that was about 80-100~km in size. This is a significantly smaller parent body than in the Themis family's case and probably about factor 2 smaller than the parent body of the Lixiaohua family. 

The largest member of the family, (778) Theobalda, has been spectroscopically classified as an F-type asteroid \citep[][]{1989aste.conf.1139T}. The data on other family members are limited, but available taxonomic classifications include C-, F-, and X-type objects, indicating their likely primitive compositions. This is further supported with an average albedo of family members of about $p_v=0.062$, based on the data of about 100 objects \citep[][]{masiero15_asteroidIV}. Apart from this limited data, Theobalda is a relatively poorly studied family, and additional observations are strongly encouraged to characterise the family members better.

Two MBCs (P/2016 J1 and 427P) are dynamically associated with the family. For both objects, sublimation of water-ice seems to be the most plausible mechanism of the mass loss. Nevertheless, both MBCs are very intriguing and deserve additional studies. The P/2016 J1 is actually a double component object, and in this respect, we can also say that there are three MBCs associated with the Theobalda family. \citet[][]{2017ApJ...837L...3M} found that P/2016 J1 split into two components (J1-A and J1-B) only several years ago, making it also the youngest known the so-called asteroid pair. Yet, the authors found that the
separation event and the present dust activity are unrelated. The 427P (P/2017 S5) is interesting from another point of view. Its time-series photometry lightcurve indicates possibly rapid rotation with a period of 1.4 hr. Though this period is yet to be confirmed, \citet[][]{2019AJ....157...54J} suggested that, if real, it may also play a role in the mass loss. The authors further suggested that such a rapid rotation could be a consequence of spin-up by sublimation torques. These findings open new questions on the relation and relative importance of sublimation and rotationally induced mass loss. For instance, one possibility is that the fast rotation is preceding the sublimation-driven activity, possibly causing surface landslides that may expose at the surface some shallow ice, which in turn triggers the sublimation driven activity. Alternatively, the sublimation-driven activity produces torques that could spin up the small asteroids to the point when they also start to shed some mass due to the fast rotation or even disintegrate into smaller fragments.

\begin{figure*}
 \begin{center}
  \includegraphics[width=0.99\textwidth, angle=-90]{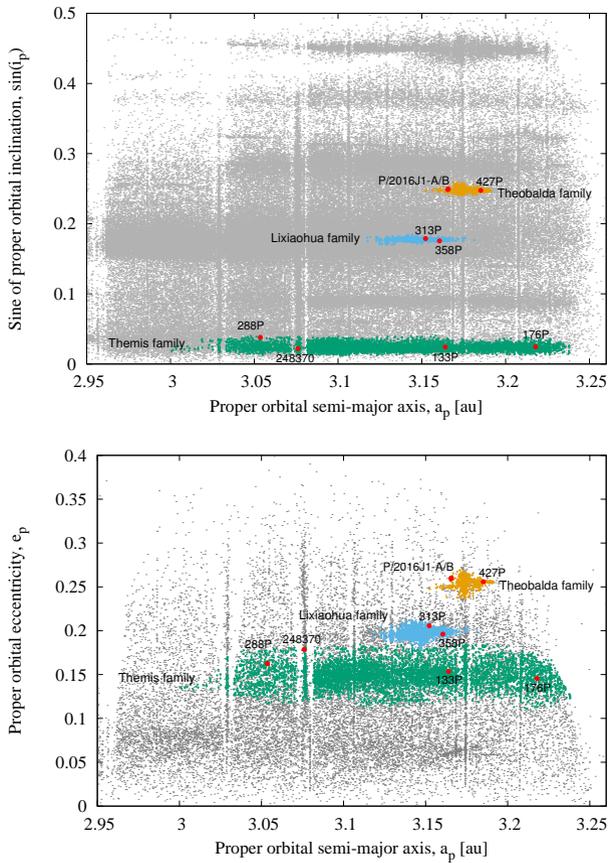}
 \end{center}
 \caption{Members of three outer main-belt asteroid families containing at least two main-belt comets projected into the space of proper orbital elements. The locations of associated MBCs are also indicated with red circles. The background asteroids are shown as gray dots.}
\label{fig:fam_mbc}
\end{figure*}

\subsubsection{Future prospects and potentially new reservoirs of main-belt comets}
\label{sss:future:mbc}

Interestingly, the three families with the largest number of associated MBCs are of quite different ages. Even considering that 133P and 288P could be members of much younger subfamilies of the Themis family, the young age does not seem to be a key for a family to contain MBCs. Therefore, it seems that links between the families and main-belt comets are primarily based on family compositions and not on their ages. In this respect, the dark families of primitive composition are the main reservoirs of MBCs, and, therefore, could also represent large water-ice reservoirs in the main asteroid belt.

There is still a lot of work to be done to understand these links properly. \citet[][]{2020A&A...643A.102D} recently studied the diversity of physical properties among five primitive outer main-belt families: Themis, Hygiea, Ursula, Veritas, and Lixiaohua families. The authors found that members of the Themis and Hygiea families show similar hydration levels. In contrast, the Ursula and Lixiaohua families present no sign of hydrated members based on the analysis of visible spectra. The Veritas family, however, presents the highest fraction of hydrated members. These results are difficult to interpret based on the current associations of main-belt comets and families. In particular, why the Lixiaohua family, which is undoubtedly linked to some MBCs, shows no signs of hydration, while members of the Veritas family, still not linked to any MBCs, seem to be the most hydrated objects among the five studied primitive families? Certainly, it could be because many MBCs are yet to be discovered, and we might find many of them among the Veritas family members. A good example is the first association of an active asteroid, namely P/2017~S8, to the Pallas family \citep[][]{2019EPSC...13..892N}.

A potentially significant reservoir of MBCs could also be the Euphrosyne asteroid family, yet another group of primitive C-type asteroids in the outer main belt \citep[][]{2020A&A...643A..38Y}. The recent work by \citet[][]{2020A&A...641A..80Y} suggested that the largest family member, asteroid (31) Euphrosyne, probably contains a significant fraction of water ice in its interior. This further suggests that the other family members could contain some water-ice. There is, however, still no known main-belt comet belonging to this family, though P/2016~P1 (PANSTARRS) could be a member of the Euphrosyne family according to \citet[][]{2019EPSC...13..892N}.

Many new MBCs have been discovered every year, and this trend could even accelerate in the near future. This will open numerous opportunities to study further links between the families and main-belt comets.

Let us conclude this section with another interesting point of some MBCs being potentially binary asteroids. \citet[][]{2017Natur.549..357A} found that the active member of the 288P group is a binary asteroid. Additionally, one of the MBCs belonging to the Theobalda family (see below), namely P/2016 J1 (PANSTARRS), is a recently formed asteroid pair, which could also be previously a binary object. More recently, \citet[][]{2021arXiv211109900J} obtained a lightcurve of another MBC, namely 331P/Gibbs (P/2012F5), and found indications that it could be a contact binary asteroid. These results open questions about whether there is a link between the MBCs and binary asteroid formation and how both these populations are related to asteroid families. Note that a binary MBC belongs to a primitive young asteroid family in each of these three cases. On the other hand, some members of the very young Datura family could be contact binaries as well (see Section~\ref{yf}), but the Datura family is not of a primitive composition. While we do expect contact binaries to form in the process of re-accumulation of fragments following asteroid disruption \citep[][]{2020Icar..33913603C}, it remains to be explained how the MBCs fit into the more global picture.

\section{Summary and Conclusions}
\label{sec:conclusions}

Asteroid families-related topics are an emerging field of solar system research. More than a century of investigation of these ice-rocky relatives brought a lot of big discoveries and outstanding accomplishments. All of these help us better understand the families and the solar system as a whole. Nevertheless, there are still open questions and problems, and
the future is bringing new challenges. Ongoing and future ground- and space-based instruments supply new observational data that is increasing fast in qualitative and quantitive ways. The large data sets will raise additional challenges for scientists working on asteroid families. Likely, big challenges always create big opportunities, and we anticipate a bright future for asteroid families related research.

A large number-density of asteroids will require substitute techniques to identify new asteroid families or attribute recent members to the existing ones. Some promising approaches are already in place. Nevertheless, these techniques will need further development and improvement to cope with the rapid increase of new asteroid discoveries. Combining different methods such as multi-step and machine learning-based algorithms could be one promising direction to follow. In this respect, it is worth reminding that any development involving machine learning should also include steps necessary to understand the algorithms and secure the reproducibility of the results.

Age determination techniques improved significantly in the last about 15 years, primarily through the backward orbit integration methods for young families and Yarkovsky-based chronology for older ones. However, the uncertainties in their determination, especially for old families, are still often considerable, and therefore this issue needs to be addressed soon. The first step in this direction could be extensive testing of existing methods. On one side, this should allow quantifying their uncertainties better. On the other hand, testing should identify the principal sources of errors, setting foundations for further improvements in age determination methods.

Studies of older families ($>$50 Myr old) will benefit more from the new physical data and refined evolution models. Combined with more accurate ages, these will allow a more authentic and detailed reconstruction of the long-term evolution. The area of young asteroid families ($<$10 Myr old), and especially very young ones ($<$1.5 Myr), is already advancing with the discovery of smaller asteroids. Many of these groups are increasing fast in the number of known members, allowing much deeper analysis. This will be the case even more in the near future. Additionally, the number of young families is far from complete, with many of them waiting to be discovered. This will increase the statistical sample of young families and provide a view of the diversity of the population. We recently already saw an example of such progress when \citet{hob2021} found that the Hobson family could be formed from the binary parent body. Nevertheless, this is just the tip of the iceberg. Many discoveries are yet to happen.

Ultimately, revealing the secrets of asteroid families is not a single-step process but rather an accumulation of knowledge over the years. Even the best and the most remarkable results represent, in principle, only the best we can learn about the families and their asteroids at the moment. The new data and better models will certainly, in the future, push our understanding of families even beyond our current imaginations.

\backmatter


\bmhead{Acknowledgements}

BN acknowledges support by the MSCA ETN Stardust-R, Grant Agreement n. 813644 under the European Union H2020 research and innovation program, and the Ministry of Education, Science and Technological Development of the Republic of Serbia, contract No. 451-03-68/2022-14/200104. The work of DV was supported by the Czech Science Foundation (grant 21-11058S).








\nocite{label}

\bibliography{families}


\end{document}